\documentclass[twocolumn]{aastex62}

\usepackage{graphics,graphicx,xspace,natbib,amssymb}
\usepackage[caption=false]{subfig}
\usepackage{amsmath}
\usepackage{threeparttable}

\newcommand{\psb}{post-starburst galaxy\xspace}
\newcommand{\psbs}{post-starburst galaxies\xspace}

\newcommand{\squiggle}{SQuIGG$\vec{L}$E\xspace}
\newcommand{\oii}{[O\,{\sc ii}]\xspace}
\newcommand{\oiii}{[O\,{\sc iii}]\xspace}
\newcommand{\fburst}{$f_{\rm{burst}}$\xspace}
\newcommand{\tburst}{$t_{\rm{burst}}$\xspace}
\newcommand{\tq}{$t_q$\xspace}
\newcommand{\comment}[1]{}

\shorttitle{\squiggle}
\shortauthors{Suess et al.}

\begin{document}

\title{\squiggle: Studying Quenching in Intermediate-$\lowercase{z}$ Galaxies--- Gas, Angu$\vec{L}$ar Momentum, and Evolution}

\author{Katherine A. Suess}
\affiliation{Astronomy Department, University of California, Berkeley, CA 94720, USA} 
\affiliation{Department of Astronomy and Astrophysics, University of California, Santa Cruz, 1156 High Street, Santa Cruz, CA 95064 USA}
\affiliation{Kavli Institute for Particle Astrophysics and Cosmology and Department of Physics, Stanford University, Stanford, CA 94305, USA}

\author{Mariska Kriek} 
\affiliation{Astronomy Department, University of California, Berkeley, CA 94720, USA}
\affiliation{Leiden Observatory, Leiden University, P.O.Box 9513, NL-2300 AA Leiden, The Netherlands }

\author{Rachel Bezanson}
\affiliation{Department of Physics and Astronomy and PITT PACC, University of Pittsburgh, Pittsburgh, PA, 15260, USA} 

\author{Jenny E. Greene}
\affiliation{Department of Astrophysical Sciences, Princeton University, Princeton, NJ 08544, USA}

\author{David Setton}
\affiliation{Department of Physics and Astronomy and PITT PACC, University of Pittsburgh, Pittsburgh, PA, 15260, USA} 

\author{Justin S. Spilker}
\altaffiliation{NHFP Hubble Fellow}
\affiliation{Department of Astronomy, University of Texas at Austin, 2515 Speedway, Stop C1400, Austin, TX 78712, USA}

\author{Robert Feldmann}
\affiliation{Institute for Computational Science, University of Zurich, CH-8057 Zurich, Switzerland}

\author{Andy D. Goulding}
\affiliation{Department of Astrophysical Sciences, Princeton University, Princeton, NJ 08544, USA}

\author{Benjamin D. Johnson}
\affiliation{Center for Astrophysics | Harvard \& Smithsonian, 60 Garden Street, Cambridge, MA 02138, USA}

\author{Joel Leja}
\affiliation{Department of Astronomy \& Astrophysics, The Pennsylvania State University, University Park, PA 16802, USA}
\affiliation{Institute for Computational \& Data Sciences, The Pennsylvania State University, University Park, PA, USA}
\affiliation{Institute for Gravitation and the Cosmos, The Pennsylvania State University, University Park, PA 16802, USA}

\author{Desika Narayanan}
\affiliation{Department of Astronomy, University of Florida, 211 Bryant Space Science Center, Gainesville, FL, 32611, USA}
\affiliation{University of Florida Informatics Institute, 432 Newell Drive, CISE Bldg E251 Gainesville, FL, 32611, US}
\affiliation{Cosmic Dawn Centre at the Niels Bohr Institue, University of Copenhagen and DTU-Space, Technical University of Denmark}

\author{Khalil Hall-Hooper}
\affiliation{Department of Mathematics, North Carolina State University, 2108 SAS Hall Box 8205, Raleigh, NC 27695, USA}

\author{Qiana Hunt} 
\affiliation{Department of Astronomy, University of Michigan, 1085 S University, Ann Arbor, MI 48109, USA}

\author{Sidney Lower}
\affiliation{Department of Astronomy, University of Florida, 211 Bryant Space Science Center, Gainesville, FL, 32611, USA}

\author{Margaret Verrico}
\affiliation{Department of Physics and Astronomy and PITT PACC, University of Pittsburgh, Pittsburgh, PA, 15260, USA}

\email{suess@ucsc.edu}

\begin{abstract}
We describe the \squiggle survey of intermediate-redshift post-starburst galaxies. We leverage the large sky coverage of the SDSS to select ${\sim}1300$ recently-quenched galaxies at $0.5<z\le0.9$ based on their unique spectral shapes. These bright, intermediate-redshift galaxies are ideal laboratories to study the physics responsible for the rapid quenching of star formation: they are distant enough to be useful analogs for high-redshift quenching galaxies, but low enough redshift that multi-wavelength follow-up observations are feasible with modest telescope investments. We use the \texttt{Prospector} code to infer the stellar population properties and non-parametric star formation histories of all galaxies in the sample. We find that \squiggle galaxies are both very massive ($\rm{M}_* \sim 10^{11.25}\rm{M}_\odot$) and quenched, with inferred star formation rates $\lesssim1\rm{M}_\odot\rm{ yr}^{-1}$, more than an order of magnitude below the star-forming main sequence. The best-fit star formation histories confirm that these galaxies recently quenched a major burst of star formation: $>75\%$ of \squiggle galaxies formed at least a quarter of their total stellar mass in the recent burst, which ended just $\sim200$~Myr before observation. We find that \squiggle galaxies are on average younger and more burst-dominated than most other $z\lesssim1$ \psb samples. 
This large sample of bright post-starburst galaxies at intermediate redshift opens a wide range of studies into the quenching process. In particular, the full \squiggle survey will investigate the molecular gas reservoirs, morphologies, kinematics, resolved stellar populations, AGN incidence, and infrared properties of this unique sample of galaxies in order to place definitive constraints on the quenching process.  
\end{abstract}


\section{Introduction}

For nearly a century, astronomers have divided galaxies into two distinct categories based on their morphologies: spirals and ellipticals \citep{hubble26}. This ``galaxy bimodality" has since been shown to extend to a myriad of other properties, including color, size, environment, molecular gas content, and star formation rate \citep[e.g.,][]{blanton03,kauffmann03,shen03,noeske07,wuyts11}. Star-forming galaxies tend to be large, blue, rotationally-supported, and gas-rich; in contrast, quiescent galaxies are red, smaller at fixed mass, supported by random motions, and gas-poor. However, the fundamental question of {\it why} this galaxy bimodality exists remains unanswered to the present day. What physical processes are responsible for ``quenching" star formation in galaxies? And why does a cessation of star formation appear to go hand-in-hand with structural and kinematic changes?

A variety of theoretical mechanisms have been proposed to form massive compact quiescent galaxies. These proposed mechanisms range from intense centrally-concentrated starbursts triggered by gas-rich major mergers (e.g., Hopkins et al.~2008, Wellons et al.~2015), very early assembly in a much denser universe (e.g., Naab et al.~2009, Wellons et al.~2015), or ``morphological" quenching, where the transition from disk to spheroid stabilizes gas reservoirs \citep[e.g.,][]{martig09}. 
While all of these mechanisms could shut down star formation by depleting, removing, heating, or stabilizing molecular gas reservoirs, the resulting quenched galaxies differ in key observables including age gradients, rotational support, and morphology. Detailed multi-wavelength follow up of quenched galaxies holds the key to distinguishing between these theoretical quenching mechanisms. 
This multi-wavelength follow-up is most effective when performed in galaxies that have {\it just} concluded their major star-forming phase: the signatures of the quenching mechanism should still be apparent, and not diluted by later mergers or secular evolution. The question then becomes how to identify these recently-quenched galaxies. 

Growing observational evidence suggests that there are at least two distinct pathways to quenching: galaxies can quench rapidly through the post-starburst phase, or quench slowly through the green valley phase \citep[e.g.,][]{barro13,barro14,schawinski14,wild16,carnall18,forrest18,wu18,rowlands18,belli19,suess21}. In this paper, we focus on the rapid quenching mode. These fast-quenching post-starburst galaxies, sometimes also called ``K+A" or ``E+A" galaxies, experienced a recent burst of intense star formation that concluded within the past $\sim$gigayear (e.g., \citealt{dressler83,couch87}; see \citealt{french21} for a recent review). Observationally, these galaxies are characterized by spectra that are dominated by A-type stars: the burst finished long enough ago that the most massive stars have died, but recently enough that the slightly longer-lived and less massive stars are still alive to dominate the optical spectrum. These spectra typically show strong Balmer breaks, deep Balmer absorption lines, weak or absent nebular emission lines, and blue slopes redward of the Balmer break. 

A variety of observational techniques have been used to select \psbs over a wide redshift range. ``E+A" or ``K+A" post-starburst galaxies in the local Universe are often selected by their high H$\delta$ equivalent widths and low [OII] luminosities; in combination, these spectral features indicate significant recent star formation but little ongoing star formation \citep[e.g.,][]{zabludoff96,dressler99,goto05,brown09,french15}. Some works use spectral template fitting approaches, which can identify galaxies whose light is dominated by young stars \citep[e.g.,][]{quintero04,pattarakijwanich16}. An alternate technique pioneered by \citet{wild14} uses principal component analysis (PCA), which effectively selects for the unique spectral shape of \psbs 
\citep[see also, e.g.,][]{almaini17,maltby18}. A similar spectral shape indentification approach was also used by \citet{kriek10}, using synthetic rest-frame colors instead of PCA-based ``supercolors". Finally, some studies have selected \psbs based on their location in the $UVJ$ plane (e.g., \citealt{whitaker12_psb,belli19,suess20}; see also \citealt{akins21}). These different methods of selecting \psbs produce samples that are relatively similar, but do not fully overlap; we will explore the differences in these sample selection methods in more detail later in this paper. 

While massive \psbs are relatively rare across cosmic time \citep[making up $\lesssim5\%$ of the total $\rm{M}_* > 10^{10}\rm{M}_\odot$ population at $z\le2$,][]{wild16}, their number density evolves rapidly with redshift. 
At $z\sim2$, \psbs are common enough to explain half of the total growth in the red sequence; by $z\sim1.4$, they represent just $\sim20\%$ of all transitioning galaxies \citep{belli19}. By $z\sim0$, massive \psbs are so rare that they contribute negligibly to the growth of the quiescent population \citep{rowlands18}. These results imply that, while \psbs play an important role in quenching at cosmic noon, their importance to the overall landscape of galaxy transformation diminishes towards lower redshifts.

The nature of \psbs may also differ between $z\sim0$ and $z\sim1$. Higher-redshift \psbs generally appear to be more burst-dominated than their low-redshift counterparts: simple modeling of H$\delta$ and $D_n4000$ values indicates that $z\sim0$ \psbs formed just $5-10\%$ of their mass in the recent burst, whereas $z>0.5$ \psbs appear to have formed the majority of their mass in the recent burst \citep{suess17}. Full spectral modeling of post-starburst star formation histories (SFHs) confirms these findings: the median burst mass fraction of $z\sim0$ \psbs is $\sim10\%$ \citep{french18}, in contrast to the $\sim70\%$ median burst mass fractions of $z\sim1$ \psbs \citep{wild20}. Low-redshift \psbs also tend to be less massive than their higher-redshift counterparts (e.g., \citealt{wild16,rowlands18}), although they are not so low-mass that environmental effects play significant roles in their quenching, (e.g. \citealt{zabludoff96}; \citealt{feldmann11}). Taken together, these results indicate that high-redshift \psbs are in the process of rapidly quenching their {\it primary} epoch of star formation, whereas low-redshift \psbs are rapidly shutting down a smaller burst of late-time star formation that does not contribute as significantly to their stellar mass. 

This redshift dependence complicates observational studies of quenching: pinpointing the physics responsible for shutting down the bulk of star formation requires looking beyond the local universe, but detailed follow-up observations require significant (and potentially prohibitive) telescope investments at $z\gtrsim1$. Even spectroscopic confirmation of \psbs is difficult at $z>1$, with samples consisting of a few tens of galaxies \citep[e.g.,][]{bezanson13,maltby16,wild20} as opposed to the hundreds or thousands of $z\lesssim0.1$ \psbs that can be identified from large all-sky surveys \citep[e.g.,][]{dressler99,quintero04,goto05,brown09,french15,alatalo16}.

Our goal in this paper is to bridge the gap between these low- and high-redshift samples by selecting bright {\it intermediate-redshift} \psbs at $0.5<z\lesssim1$ that will serve as the ideal testbeds to understand the physics of quenching. We aim to find a large sample of galaxies that have high burst mass fractions and are therefore in the process of shutting down their major epoch of star formation. At the same time, we want these galaxies to be bright enough to conduct full multi-wavelength follow-up studies with reasonable telescope investments. These bright, massive, intermediate-redshift \psbs will not provide us with a complete census of how star formation shuts down across all stellar masses and timescales; however, they will serve as excellent laboratories to study how rapid quenching proceeds in massive galaxies. This sample selection is now possible due to the SDSS, which provides millions of spectra over our targeted redshift range. This enormous public database allows us to identify a statistically large sample of rare bright, young post-starburst systems. 
With this sample, we will be Studying Quenching in Intermediate-z Galaxies-- Gas, angu$\vec{L}$ar momentum, and Evolution (\squiggle).
The \squiggle sample will serve as the ideal testbed to study the quenching process in massive galaxies in detail. 

This paper is organized as follows. In Section~\ref{sec:selection}, we describe our post-starburst sample selection. We also generate mock galaxy spectra with known properties, and use those mocks to investigate what types of galaxies are included in our sample selection. In Section~\ref{sec:prospector}, we use the \texttt{Prospector} Bayesian stellar population synthesis code to fit the stellar masses, dust properties, and star formation histories of the galaxies in our sample. In particular, we use these derived star formation histories to investigate how long \squiggle \psbs have been quiescent, the fraction of their mass that was formed in a recent starburst, and their ongoing star formation rates (SFRs). In Section~\ref{sec:line_sfrs}, we calculate SFRs from spectral lines present in the SDSS spectra of our galaxies, and compare these results to the \texttt{Prospector} SFRs. In Section~\ref{sec:comparison}, we place the \squiggle sample in context by comparing it to other post-starburst samples. Finally, in Section~\ref{sec:objectives} we briefly describe the science objectives of the remainder of the \squiggle survey. Throughout this paper we assume a flat $\Lambda$CDM cosmology with $\Omega_{\rm m}=0.3$, $\Omega_\Lambda=0.7$, and $h=0.7$. We also assume a \citet{chabrier03} initial mass function.

\section{Sample Selection}
\label{sec:selection}
In this section, we describe how we select \psbs from the SDSS. Our selection is designed to select the brightest, most massive, most burst-dominated \psbs at $z\sim0.7$, with the goal of identifying galaxies that are ideal targets for multi-wavelength follow-up studies of the quenching process. 
We explore the differences between our sample selection algorithm and those used in previous studies in Section~\ref{sec:comparison}.

\begin{figure}
    \centering
    \includegraphics[width=.48\textwidth]{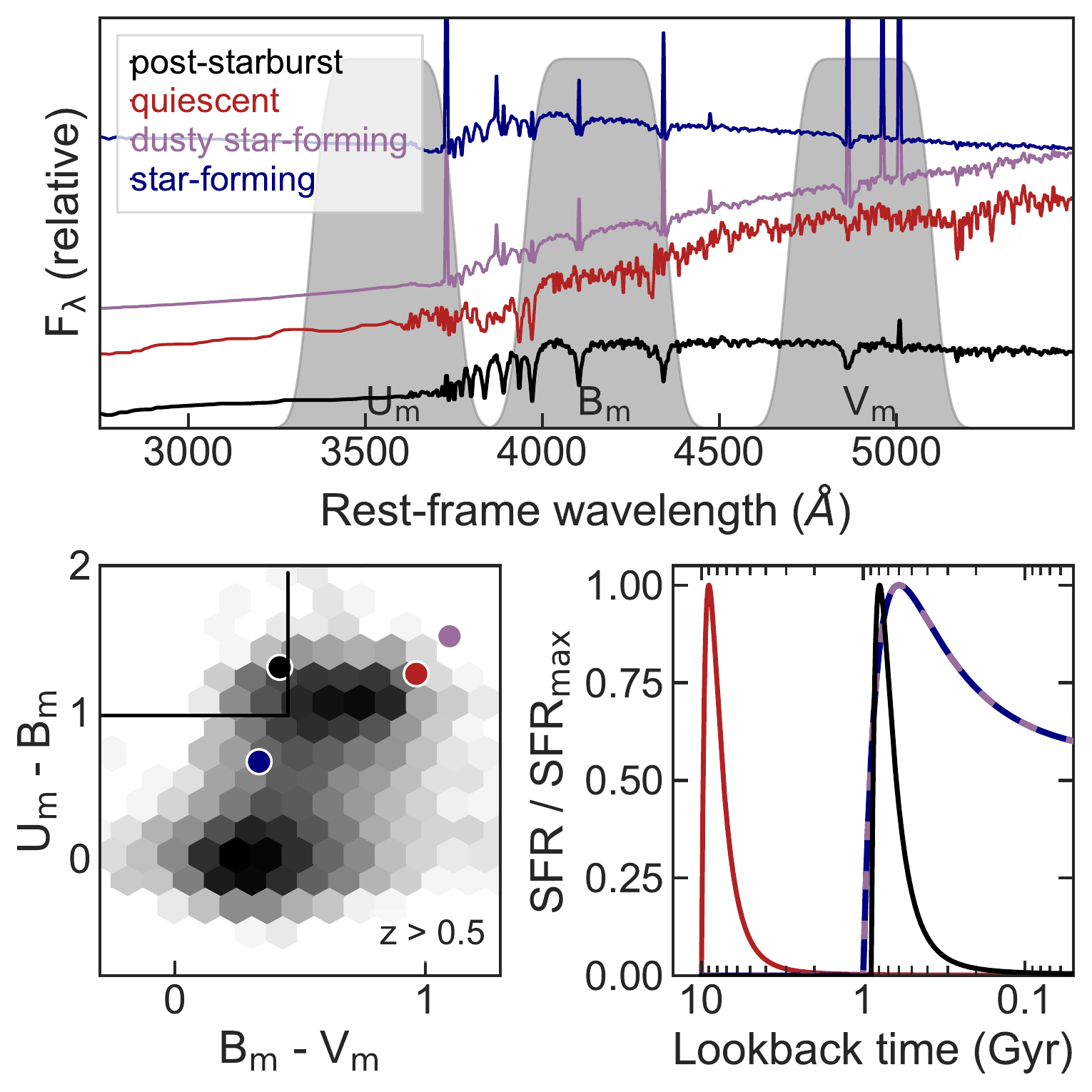}
    \caption{Demonstration of color-based selection technique. Top: grey shaded regions show the synthetic rest-frame $U_m$, $B_m$, and $V_m$ filters. Blue, purple, red, and black curves show example synthetic spectra of an unobscured star-forming galaxy, a dusty star-forming galaxy, a quiescent galaxy, and a post-starburst galaxy; spectra have an arbitrary factor added to $F_\lambda$ for visibility. Bottom left: $U_m-B_m$ as a function of $B_m-V_m$; the shaded background represents the relative density of all $z>0.5$ galaxies in the SDSS, while the black, blue, purple, and red points show the colors of the spectra shown in the top plot. The black lines show our color criteria for a galaxy to be selected as post-starburst. Post-starburst galaxies can be distinguished from other galaxy types in this color-color space: the $U_m-B_m$ color separates \psbs from unobscured star-forming galaxies, while the $B_m-V_m$ color separates \psbs from dusty star-forming galaxies and older quiescent galaxies. Bottom right: Star formation histories of the four synthetic spectra shown in the top plot. The post-starburst galaxy (black) has a clearly different SFH than either the quiescent galaxy, which formed its stars much earlier, or the two star-forming galaxies, which have significant ongoing star formation.}
    \label{fig:selection}
\end{figure}

For the \squiggle sample, we follow the \citet{kriek10} method and select \psbs from the SDSS based on their unique spectral shapes. Synthetic medium-band rest-frame $U_m$, $V_m$, and $B_m$ filters are designed to target both the Balmer break and the slope of the spectrum redward of the break; as shown in Figure~\ref{fig:selection}, these two rest-frame colors alone are effective at selecting \psbs.   
Model spectra \citep[generated with FSPS,][]{conroy09,conroy10} of a post-starburst galaxy (black), a fully quiescent galaxy (red), an unobsured star-forming galaxy (blue) and a dusty star-forming galaxy (purple) are shown over the same wavelength range. 
The bottom left panel in Figure~\ref{fig:selection} shows the location of these four spectra in $UBV$ space, as well as the density of all SDSS spectra with $z>0.5$. Our post-starburst color selection cuts, described in detail below, are shown with black lines. The color cuts include the post-starburst galaxy while excluding both star-forming and older quiescent galaxies; just $\sim5\%$ of all $z>0.5$ galaxies in the SDSS satisfy these color cuts. 
Figure~\ref{fig:selection} shows that unobscured star-forming galaxies--- including galaxies which may have ongoing starbursts--- have similar $B_m-V_m$ colors to \psbs, but weaker Balmer breaks that do not satisfy our  $U_m-B_m$ cut. In contrast, both dusty star-forming galaxies and older quiescent galaxies have similar $U_m-B_m$ colors as \psbs, but redder $B_m-V_m$ colors. 
Spectrally, older quiescent galaxies and post-starburst galaxies can be distinguished by the differences in the strength of the Balmer break, the slope of the spectrum redward of the break, and the depth of both the Balmer absorption lines and the Calcium H \& K lines. Dusty star-forming galaxies differ from post-starburst galaxies mainly in the slope of the spectrum redward of the Balmer break and the strength of the nebular emission lines.

Our parent sample consists of all 1,921,000 galaxies in the SDSS DR14 spectroscopic catalog \citep{sdss_dr14} with $z>0.5$. We then make a quality cut which removes non-physical spectra: we calculate the $r-i$ and $i-z$ colors from the SDSS spectrum, and remove any galaxies for which these spectral colors differ from the SDSS photometric colors by more than 0.25~dex. 
We note that at these redshifts, the fiber encompasses $\gtrsim75\%$ of the total light from these compact post-starburst galaxies; typically the spectral and photometric colors are within 0.05~dex. 
We ensure that the wavelength range of the SDSS spectrum covers our rest-frame $U_m$, $B_m$, and $V_m$ filters, then calculate the flux in each synthetic rest-frame filter. To ensure high quality spectra, we require a signal-to-noise ratio (S/N) $\ge 6$ in both our $B_m$ and $V_m$ filters; this S/N cutoff was chosen by examining representative spectra by eye. We do not require a minimum S/N in the $U_m$ filter, because post-starburst galaxies typically have minimal UV flux and thus low S/N blueward of the Balmer break (see, e.g., Figure~\ref{fig:selection}). 32,168 galaxies remain in the parent sample after these initial quality cuts. Finally, we select all objects in this sample with $U-B > 0.975$ and $-0.25 < B-V < 0.45$ as post-starburst galaxies. Our final sample includes 1,318 unique post-starburst galaxies. These galaxies form the full \squiggle sample.

\begin{figure}
    \centering
    \includegraphics[width=.49\textwidth]{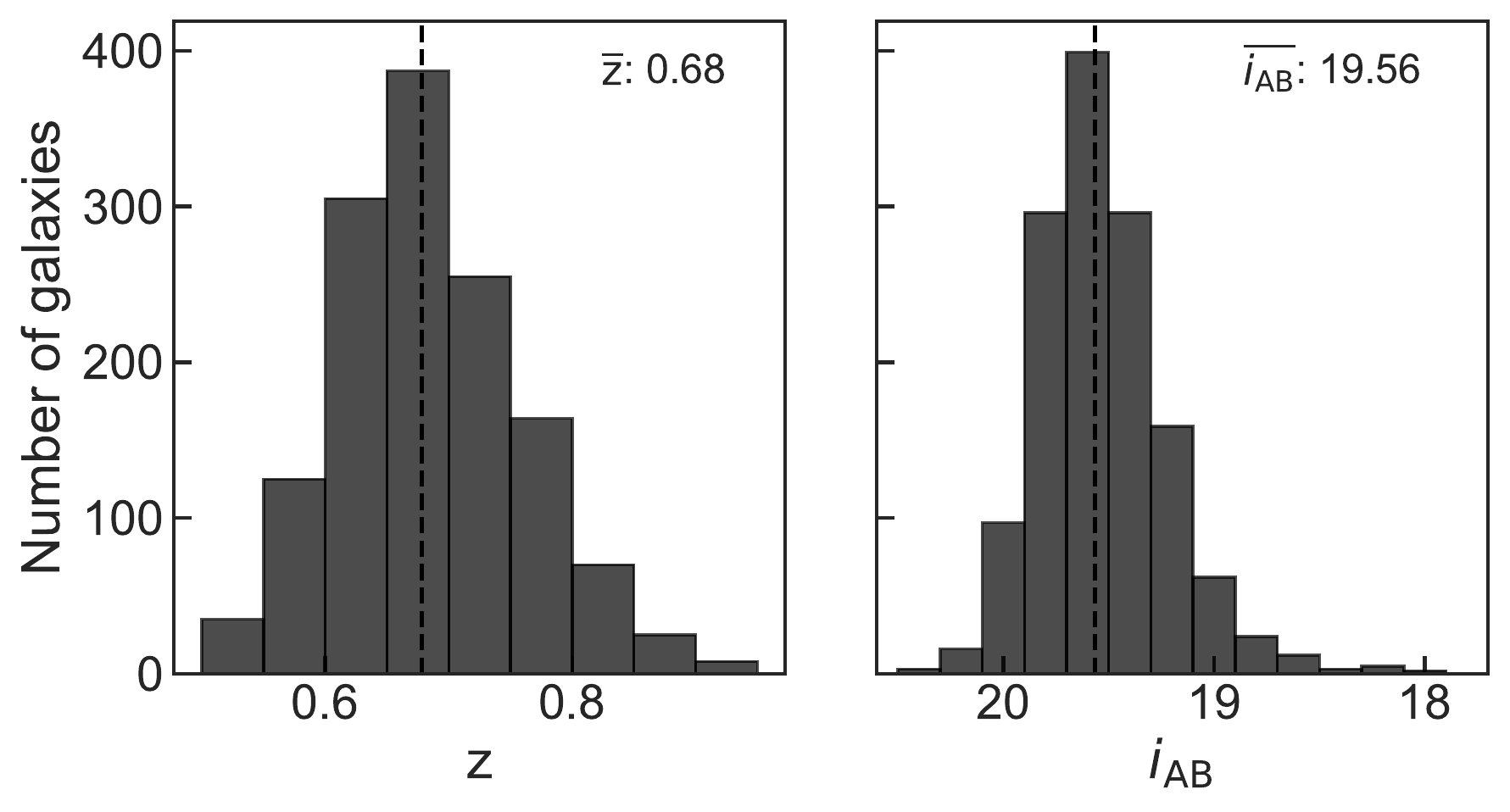}
    \caption{Histograms of the redshift and $i$-band magnitude of all \squiggle galaxies.}
    \label{fig:hist}
\end{figure}

Figure~\ref{fig:hist} shows histograms of the redshift and $i$-band magnitude distribution of the full \squiggle sample. By construction, all galaxies have $z\ge 0.5$. The median sample redshift is $z=0.68$, with a tail of galaxies up to $z=0.94$. This redshift distribution is effectively a competition between apparent magnitude (there are fewer high-S/N SDSS spectra at higher redshift) and the number density evolution of post-starburst galaxies \citep[there are many more post-starburst galaxies at higher redshift, e.g.,][]{wild16,belli19}. The $i$-band magnitudes of \squiggle galaxies range from 17.9 to 20.5, with the faint end cutoff primarily driven by our S/N cut.

\begin{figure}
    \centering
    \includegraphics[width=.49\textwidth]{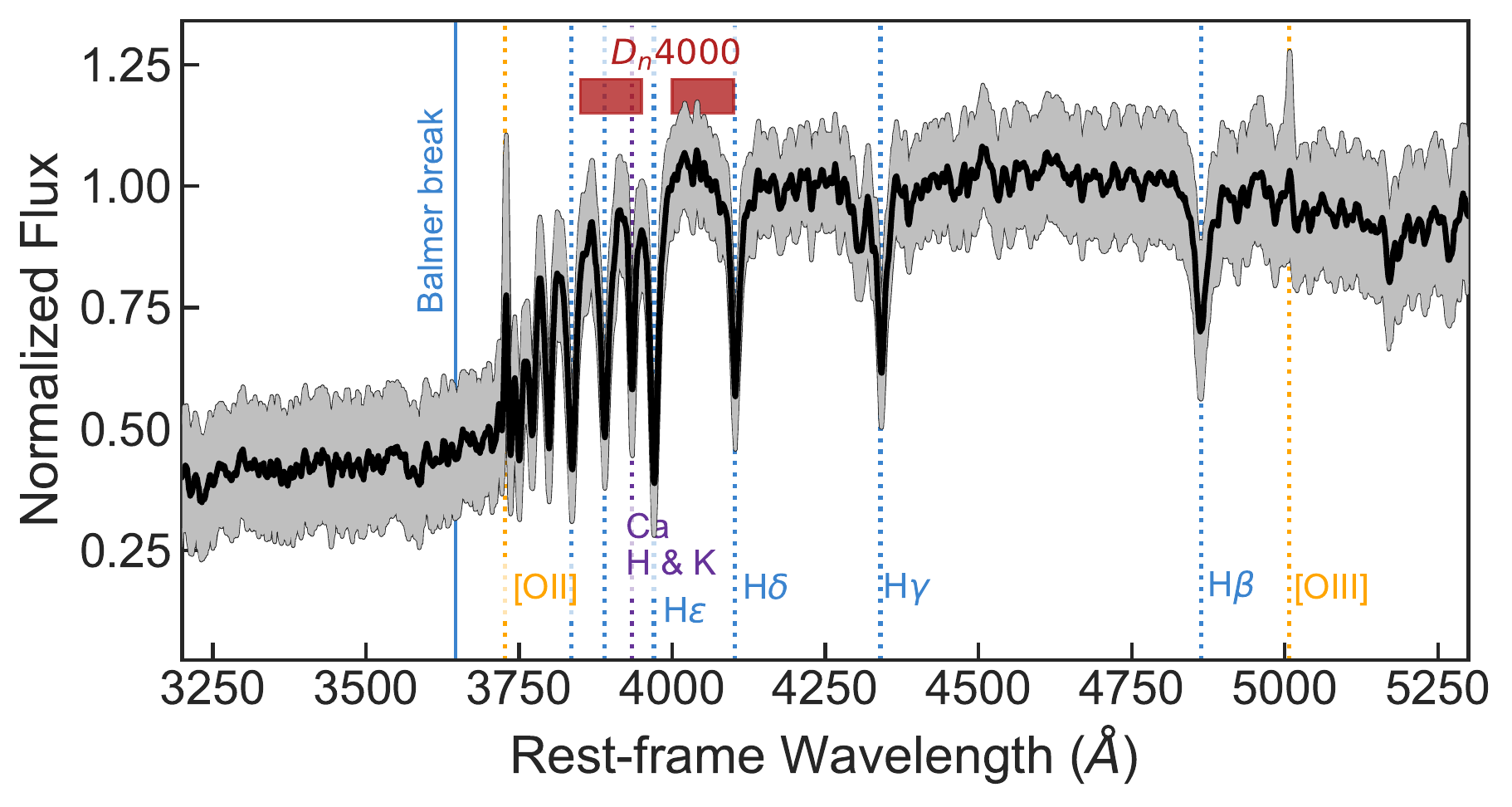}
    \caption{Stack of all post-starburst galaxies in the \squiggle sample. The black line shows the median stacked spectrum; the grey shaded region indicates the 16-84th percentile range of all spectra. \squiggle galaxies have strong Balmer breaks, blue slopes redward of the break, deep Balmer absorption lines, and typically weak or absent \oii and \oiii emission lines.}
    \label{fig:stack_spectrum}
\end{figure}

Figure~\ref{fig:stack_spectrum} shows a stacked spectrum of all 1,318 post-starburst galaxies in \squiggle, normalized using the flux between 4150 and 4250 \AA. The grey shaded region shows the 16-84th percentile of all spectra. This stacked spectrum clearly shows the representative characteristics of a post-starburst galaxy: a strong Balmer break, deep Balmer absorption lines, and weak \oii emission. 
Spectral modeling (Section~\ref{sec:prospector}) indicates that 98\% of \squiggle galaxies have star formation rates below the main sequence, and 95\% of \squiggle galaxies formed $>10\%$ of their stellar mass in a recent burst; this indicates that our sample selection technique is very effective at identifying recently-quenched galaxies.

\begin{figure}[ht]
    \centering
    \includegraphics[width=.49\textwidth]{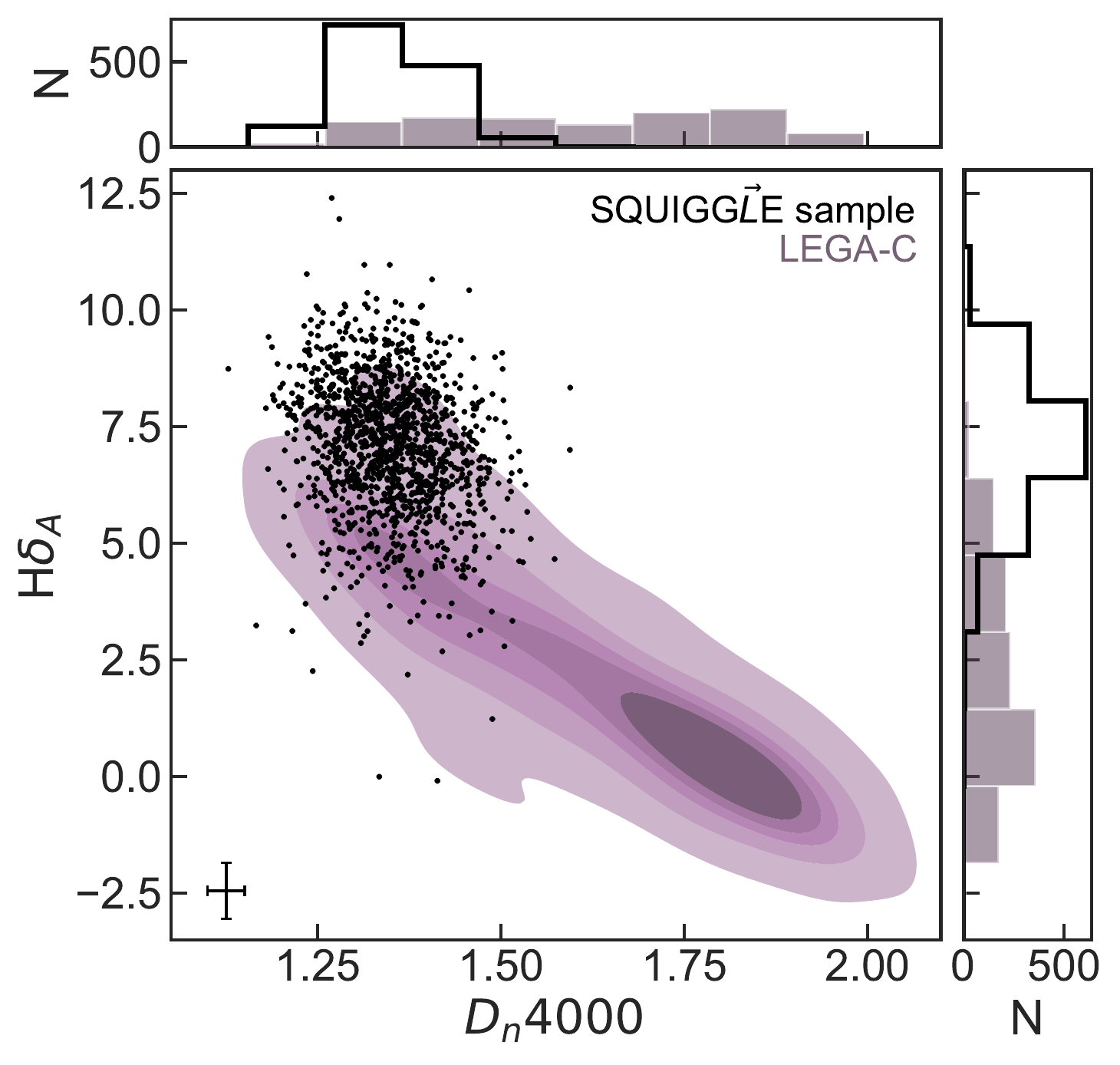}
    \caption{$H\delta_A$ as a function of $D_n4000$ for all \squiggle galaxies (black points). A characteristic error bar for \squiggle galaxies is shown in the lower left. Age increases towards the lower right of the diagram. The distribution of galaxies with similar stellar mass and redshift from the LEGA-C survey is shown by the shaded purple background. Our selection has clearly resulted in a sample of galaxies with high $H\delta$ equivalent widths and low $D_n4000$ indices, despite not explicitly selecting on these parameters.}
    \label{fig:Hd-D4000}
\end{figure}

We also measure the H$\delta$ equivalent width and the $D_n4000$ index for all galaxies in our sample. H$\delta$ traces recent star formation, while $D_n4000$ probes the age of the stellar population; together, these indices provide a fairly reliable indication of the galaxy's evolutionary stage. We use the \texttt{pyphot} python package to calculate H$\delta_A$, and obtain error bars via bootstrap resampling. We measure $D_n4000$ from the SDSS spectra using the bandpass definitions from \citet{balogh99}. In Figure~\ref{fig:Hd-D4000}, we show H$\delta_A$ as a function of $D_n4000$ for all galaxies in our sample (black points). To help place \squiggle galaxies in context, Figure~\ref{fig:Hd-D4000} also shows star-forming and quiescent galaxies from the LEGA-C spectroscopic survey \citep{vanderwel16} with $0.5\le z\le1$ and $\log{\rm{M}_* / \rm{M}_\odot}\ge 10.7$. This mass and redshift range roughly matches that of \squiggle, though LEGA-C's relatively small area means that it lacks the highest stellar masses found in \squiggle. 

Despite the fact that we do not explicitly select \psbs using either of these two indices (unlike some \psb selection techniques, which use an H$\delta_A$ cut), we see that \squiggle galaxies are clustered at high H$\delta_A$ and low $D_n4000$. This result confirms that the galaxies in the \squiggle sample are indeed fairly young. We again see that \squiggle galaxies lie in a relatively extreme region of parameter space compared to the overall galaxy population. These extreme H$\delta_A$ and $D_n4000$ values hint at the fact that \squiggle galaxies have distinct star formation histories from the majority of galaxies at similar redshifts and stellar masses; we will explore this in greater detail in Section~\ref{sec:prospector}.

Table~\ref{table:basic_props} lists the basic properties of the \psbs identified by \squiggle.

\begin{table*}
\centering
\caption{Basic properties of \squiggle \psbs}
\label{table:basic_props}
\begin{threeparttable}
\begin{tabular}{cccccccc}
\hline \hline
SDSS ID & RA (deg) & Dec (deg) & $z_{\mathrm{spec}}$ & H$\delta_A$ (\AA) & $D_n4000$ & $\sigma_*$ (km/s)\tnote{a} & aperture correction \\ \hline
spec-6137-56270-0195 & 353.79039 & 16.10073 & 0.7473 & 7.86$\pm$0.51 & 1.12$\pm$0.02 & 186$\pm$52 & 1.50\\
spec-0978-52431-0077 & 260.01290 & 30.28743 & 0.6840 & 9.45$\pm$0.33 & 1.16$\pm$0.01 & 223$\pm$42 & 1.07\\
spec-5192-56066-0419 & 238.72249 & 38.33752 & 0.7237 & 7.49$\pm$0.62 & 1.31$\pm$0.03 & 222$\pm$123 & 1.21\\
spec-5288-55865-0858 & 132.07321 & 13.07628 & 0.5233 & 8.53$\pm$0.47 & 1.19$\pm$0.02 & 211$\pm$53 & 0.95\\
spec-4575-55590-0605 & 144.35677 & 36.50522 & 0.6206 & 8.11$\pm$0.43 & 1.30$\pm$0.02 & 167$\pm$20 & 1.39\\
spec-3817-55277-0279 & 135.90519 & 3.81953 & 0.7570 & 4.54$\pm$0.50 & 1.34$\pm$0.02 & 261$\pm$46 & 1.47\\
spec-5140-55836-0177 & 21.74532 & 14.35716 & 0.6946 & 7.87$\pm$0.53 & 1.27$\pm$0.02 & 164$\pm$43 & 1.35\\
spec-1630-54476-0502 & 53.22548 & -6.20368 & 0.5715 & 5.02$\pm$0.39 & 1.30$\pm$0.02 & 239$\pm$33 & 1.29\\
spec-3754-55488-0041 & 120.22957 & 32.94343 & 0.7037 & 5.26$\pm$0.65 & 1.40$\pm$0.03 & 265$\pm$62 & 1.64\\
spec-6649-56364-0311 & 166.58688 & 45.04543 & 0.6391 & 6.53$\pm$0.59 & 1.32$\pm$0.03 & 191$\pm$35 & 1.72\\
spec-5048-56218-0165 & 337.14303 & 10.75239 & 0.6671 & 7.82$\pm$0.71 & 1.30$\pm$0.03 & 129$\pm$65 & 1.57\\
spec-6054-56089-0547 & 225.25613 & 42.77234 & 0.6094 & 5.58$\pm$0.39 & 1.30$\pm$0.02 & 234$\pm$28 & 1.45\\
spec-4403-55536-0765 & 27.85715 & 6.27124 & 0.6698 & 5.28$\pm$0.64 & 1.21$\pm$0.02 & 232$\pm$50 & 1.72\\
spec-6032-56067-0159 & 236.68717 & 45.81196 & 0.6819 & 8.75$\pm$0.63 & 1.24$\pm$0.03 & 205$\pm$64 & 1.26\\
spec-6639-56385-0597 & 177.92877 & 43.34649 & 0.7691 & 7.64$\pm$0.73 & 1.26$\pm$0.02 & 229$\pm$47 & 1.58\\
spec-4013-55629-0073 & 228.42489 & 2.08515 & 0.7409 & 5.92$\pm$0.52 & 1.29$\pm$0.02 & 146$\pm$40 & 1.48\\
spec-5993-56070-0251 & 199.47152 & 22.03277 & 0.7208 & 7.73$\pm$0.44 & 1.21$\pm$0.02 & 218$\pm$57 & 1.68\\
spec-5014-55717-0745 & 257.42214 & 27.66418 & 0.6926 & 8.97$\pm$0.69 & 1.36$\pm$0.03 & 175$\pm$52 & 1.35\\
spec-5291-55947-0601 & 132.52735 & 11.19210 & 0.6111 & 6.98$\pm$0.57 & 1.16$\pm$0.02 & 288$\pm$21 & 1.35\\
spec-5475-56011-0379 & 222.19133 & 10.16960 & 0.6462 & 7.28$\pm$0.34 & 1.23$\pm$0.01 & 209$\pm$30 & 1.71\\
... & ... & ... & ... & ... & ... & ... \\ \hline
\end{tabular}
\begin{tablenotes}
\item[a]{From the pPXF fits described in \citet{greene20}.}
\item (This table is available in its entirety in a machine-readable form in the online journal. A portion is shown here for guidance regarding its form and content.)
\end{tablenotes}
\end{threeparttable}
\end{table*}

\subsection{Why were these galaxies targeted by SDSS?}
Here, we examine the target flags of the selected \squiggle galaxies to understand why they were included in the SDSS spectroscopic sample. The vast majority of \squiggle galaxies--- 1,132 out of 1,318, $\sim82\%$--- are part of the main ``CMASS" BOSS sample \citep{dawson13}. The CMASS selection used color-magnitude cuts designed to target massive galaxies at $0.4<z<0.7$. 
An additional 11 galaxies are part of the sparse CMASS sample, which includes fainter and bluer galaxies than the main CMASS sample. 64 galaxies are part of the ``commissioning" CMASS sample, which used slightly different color cuts than the final CMASS survey. 
The remaining 111 PSBs were targeted as part of 24 different programs within SDSS. The majority of these 111 galaxies were selected as BOSS ancillary targets because their $ugriz$ or WISE colors resembled those of high-redshift quasars or LRGs. Several others were selected because they had matches in the Chandra Source Catalog \citep{evans10} or the FIRST radio survey (\citealt{becker95}, see also \citealt{greene20}). Galaxies with non-CMASS target flags tend to be slightly brighter and lie at lower redshifts than the median of the full \squiggle sample.

\subsection{Understanding our sample selection: how do physical parameters map onto $U_m-B_m$ and $B_m-V_m$ colors?}
\label{sec:selection_explain}
In order to understand the types of galaxies that fall into our color-based selection method, we generate a set of mock SDSS-like spectra then map their physical properties onto $UBV$ color space.
We 
generate these mock galaxy spectra using FSPS \citep{conroy09,conroy10} assuming a \citet{chabrier03} IMF, the \citet{calzetti00} dust law, a total stellar mass of $10^{11.25}M_\odot$, and a velocity dispersion of 200 km/s. We vary the dust attenuation ($0 \le A_v/\rm{mag} \le 2.5$), the metallicity ($0\le \log Z/Z_\odot\le 0.5$), and the spectral S/N (10-90\% noise levels of our observed \squiggle spectra). We model the SFHs of the mock galaxies as a delayed $\tau$ component plus a recent top-hat burst; we vary the mass fraction in the recent burst ($10\% \le $~\fburst$\le 99\%$), the duration of the recent burst ($100\le$~\tburst/Myr~$\le600$), the time since quenching ($0.05\le t_q/\rm{Gyr}\le1.0$), and the star formation rate after quenching (e.g., the amount of ``frosting", $10^{-5}\le \rm{SFR_q}/M_\odot\rm{yr}^{-1}\le 30$). We explain the generation of these mock spectra in more detail in a forthcoming paper, Suess et al. in prep., which examines the best methods to recover accurate SFHs for \psbs. 

\begin{figure}[ht]
    \centering
    \includegraphics[width=.48\textwidth]{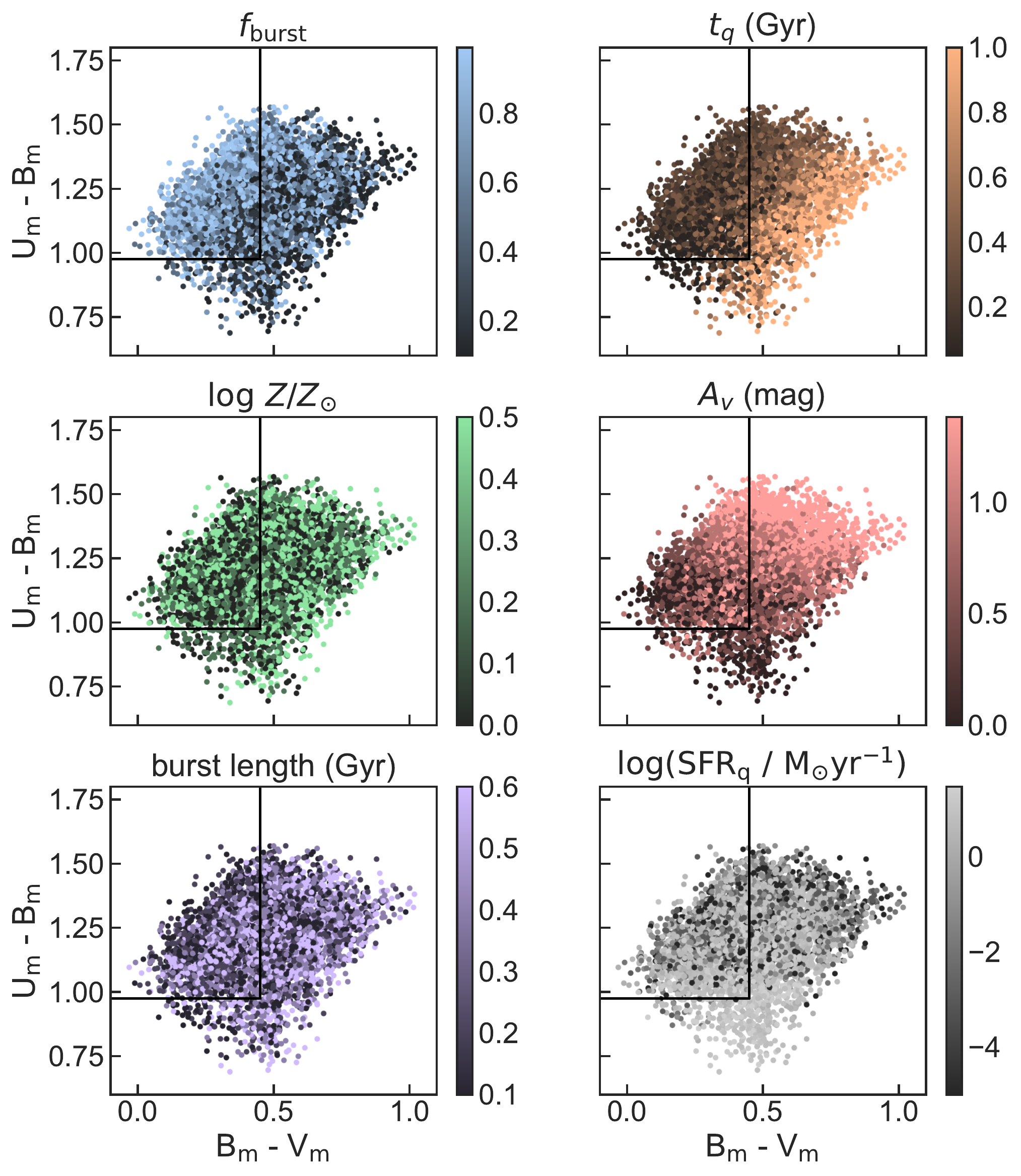}
    \caption{$U_m-B_m$ vs $B_m-V_m$ colors for 5,000 synthetic spectra generated with FSPS. The black lines show our color-based \psb sample selection. Each panel is colored by one of the parameters we vary to create the grid of synthetic spectra. $U_m-B_m$ and $B_m-V_m$ are sensitive to changes in burst mass fraction, time since quenching, dust extinction, and current SFR; however, the colors are not sensitive to metallicity or the timescale of the older burst of star formation. Galaxies that fall into our \psb selection tend to have high $f_{\rm{burst}}$, low $t_q$, low $A_v$, and relatively low SFR.} 
    \label{fig:synth_selection}
\end{figure}

We generate 5,000 mock galaxy spectra, then run our color-based selection algorithm (Section~\ref{sec:selection}) on the set of mock spectra. 1,821 of the 5,000 of the mock galaxies are classified as post-starburst by our sample selection criteria. 
Figure~\ref{fig:synth_selection} shows $U_m-B_m$ versus $B_m-V_m$ for all 5,000 mock spectra. Each panel is colored by a different physical parameter, and our post-starburst color cuts are shown by the solid black lines. While \fburst, \tq, A$_v$, and SFR$_q$ show clear gradients across $UBV$ space, neither metallicity nor the length of the recent burst exhibit coherent trends. This indicates that our sample selection does not prefer galaxies of a specific \tburst or metallicity. However, the galaxies that we select as post-starburst tend to have high \fburst, low \tq, and relatively low A$_v$ and SFR$_q$. This indicates that our sample selection algorithm is sensitive to relatively dust-free galaxies that recently quenched after a large starburst--- exactly the types of galaxies we were attempting to target.

\begin{figure}
    \centering
    \includegraphics[width=.48\textwidth]{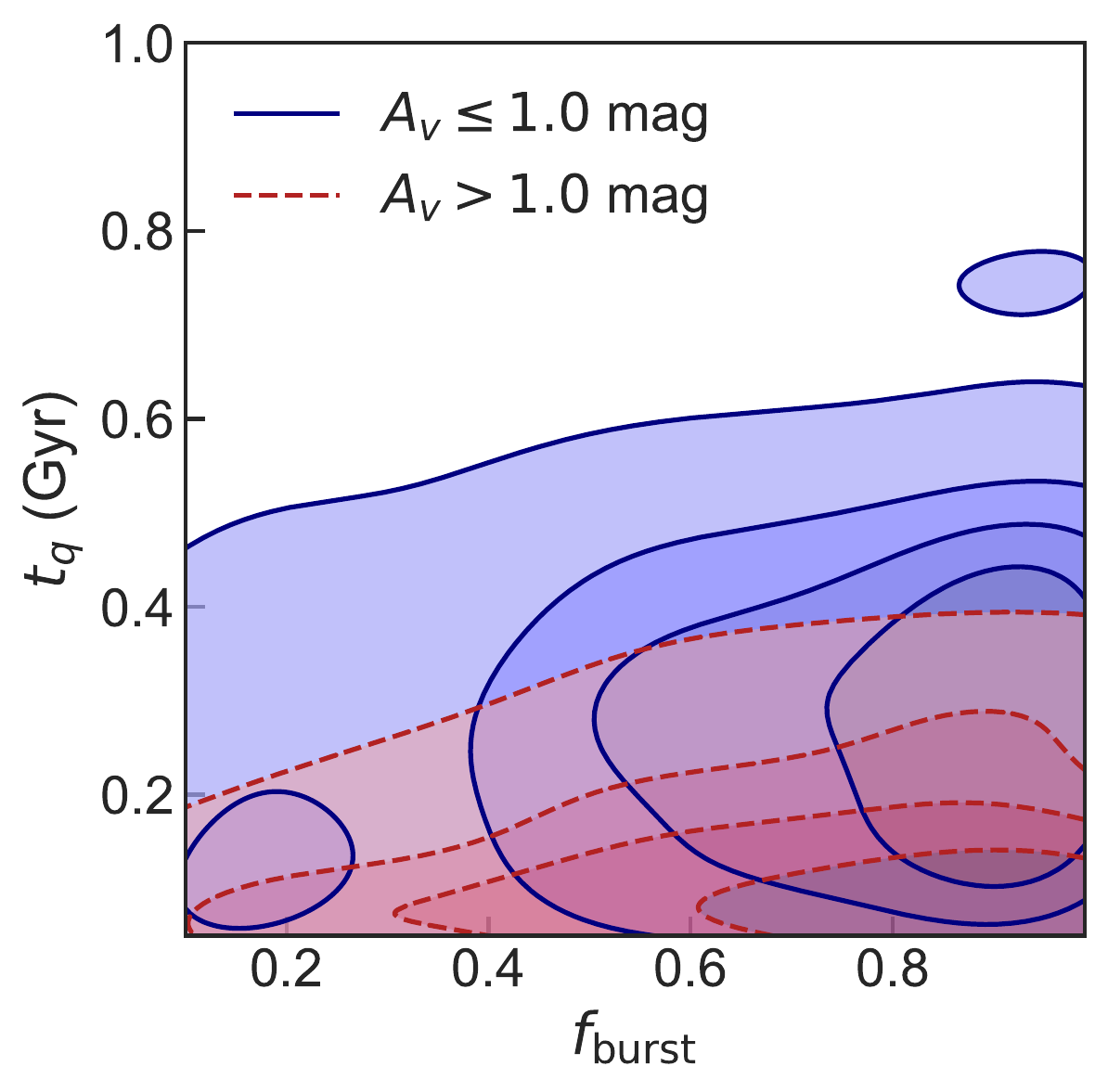}
    \caption{Time since quenching as a function of burst mass fraction for mock \psbs that fall within our selection, split into more dusty (red, A$_v>1.0$~mag) and less obscured (blue, A$_v\lesssim1.0$~mag) galaxies. 
    Our sample tends to select \psbs with relatively high \fburst and low \tq. Dusty galaxies only fall in our sample if they quenched very recently.}
    \label{fig:selection_correlations}
\end{figure}

We also briefly explore correlations between these parameters. Figure~\ref{fig:selection_correlations} shows time since quenching as a function of burst mass fraction for the mock galaxies which are classified as post-starburst by our selection algorithm. The blue contours show relatively dust-free mock \psbs, and the red contours show the dusty mock \psbs. Relatively dust-free \psbs can be found at a range of \tq and \fburst values. However, dusty galaxies are only classified as post-starburst if they quenched very recently, within the past $\sim200$~Myr. We can therefore expect that, while the majority of galaxies in the \squiggle sample are likely not highly dust-obscured, some especially young dusty galaxies may be included in the sample. We do not find a significant difference in the \tq and \fburst values of low- and high-SFR mock \psbs.

\section{extracting star formation histories via spectral energy distribution fitting}
\label{sec:prospector}

In this section, we use the \texttt{Prospector} stellar population synthesis fitting code \citep{johnson17, leja17, johnson20} to investigate the stellar masses, dust properties, metallicities, SFRs, and star formation histories (SFHs) of galaxies in the \squiggle sample. 

Robust SFHs provide a wealth of information, allowing us to investigate how long these galaxies have been quenched (\tq), the fraction of their mass that was formed in the recent burst (\fburst), and the timescale of both the recent burst and the quenching process. However, traditional ``parametric" SFH models that depend on just a few parameters impose strong priors on sSFRs and mass-weighted ages, and results from these parametric fits may not accurately reflect the true mass assembly histories of galaxies \citep[e.g.,][]{carnall19}. Previous studies have worked to mitigate these biases by modeling post-starburst SFHs as the sum of two parametric components, one for the recent burst and one for the older stellar population \citep[e.g.,][]{kaviraj07,french18,wild20}. While this approach improves on traditional parametric approaches by allowing the mass fraction in the recent burst to vary, it still explicitly imposes a specific shape for the recent burst and may thus bias results. Here, we use a ``non-parametric" form for the SFH, which allows for arbitrary SFR in adjacent timebins. This approach introduces a larger number of free parameters into the fit in exchange for allowing more flexibility and freedom in the derived SFH. These non-parametric SFHs have been shown to more accurately recover galaxy properties such as stellar mass \citep[e.g.,][]{lower20}. Here, we develop and use an non-parametric model specifically tuned to recover the SFHs of \psbs. With these fits, we allow for full flexibility in the burst shape, burst duration, burst mass fraction, and quenching timescale of \squiggle \psbs.

\subsection{SED fitting setup}
\label{sec:prospector_setup}
We fit the SEDs of all \squiggle galaxies using the \texttt{Prospector} stellar population synthesis fitting software \citep{johnson17, leja17, johnson20}. We use the \texttt{dynesty} dynamic nested sampling package \citep{speagle20}, the FSPS stellar population synthesis models \citep{conroy09,conroy10}, the MILES spectral library \citep{sanchez06,falcon11}, and the MIST isochrones (\citealt{choi16,dotter16}; based on MESA, \citealt{paxton11,paxton13,paxton15}). We assume the \citet{chabrier03} initial mass function, fix the model redshift to the SDSS spectroscopic redshift, and add nebular emission to the spectra using the default fixed parameters in \texttt{Prospector} \citep[see][]{byler17}. We fit for stellar mass and metallicity using the mass-metallicity prior described in \citet{leja19b}. We also fit for the velocity dispersion of the SDSS spectra, using a gaussian prior with the mean and sigma of the output pPXF \citep{cappellari2017} velocity dispersion fits \citep[see][]{greene20}. We assume the \citet{kriek13} dust law, with a free A$_v$ and dust index; in this dust law, the bump strength is tied to the slope. Following e.g.  \citet{wild20}, we assume that the attenuation is doubled around young ($<10^7$~yr) stars. We fix the shape of the IR SED following the \citet{draine07} dust emission templates, with $U_{\rm{min}} = 1.0$, $\gamma_e=0.01$, and $q_{\rm{PAH}}=2.0$. We also include both a spectroscopic jitter term and the \texttt{Prospector} pixel outlier model, which are designed to prevent mis-estimated spectroscopic uncertainties or bad spectral pixels from skewing the output.

\begin{figure*}[ht]
    \centering
    \includegraphics[width=.8\textwidth]{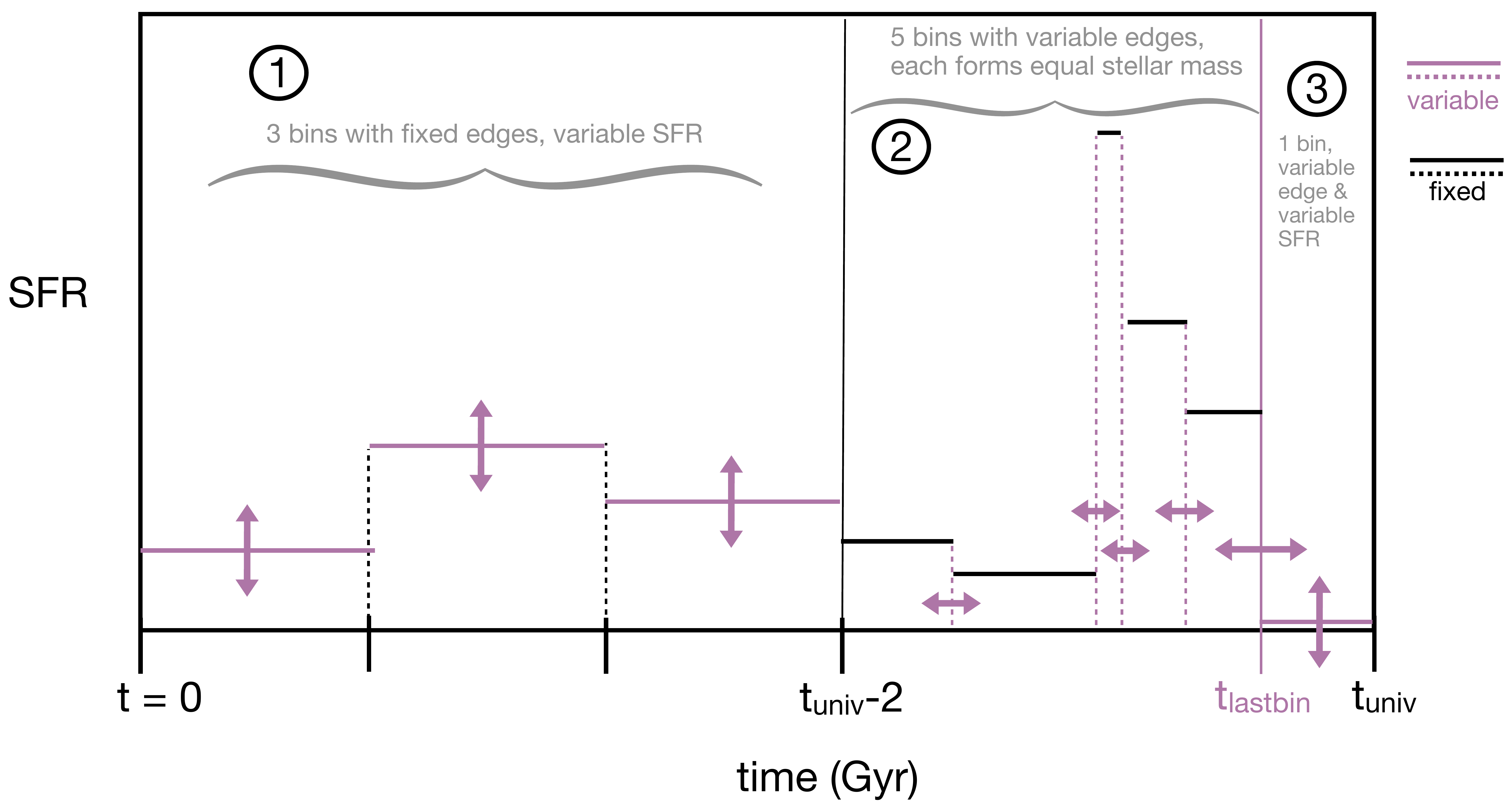}
    \caption{Graphical representation of the SFH model used in this work. The model consists of three parts: the oldest section contains three bins with variable SFR but fixed edges, the middle section contains 5 bins with variable edges, and the most recent portion is modeled as a single bin with free width and SFR. Purple lines with arrows are used to indicate quantities that are allowed to vary; black lines indicate fixed quantities.}
    \label{fig:sfh}
\end{figure*}

Our adopted SFH is a modified version of the flexible continuity prior from \citet{leja19a}, and is illustrated in Figure~\ref{fig:sfh}. The SFH includes three fixed-edge time bins at the beginning of the universe, five flexible-edge time bins covering the 2~Gyr before observation, and a final bin where we fit for both the bin length and the SFR. All five of the flexible-edge bins form an equal stellar mass, and the edges of the flexible bins are adjusted at each likelihood call based on the log(SFR) ratio between adjacent bins; this parameterization is described in detail in \citet{leja19b}. The SFH thus has nine free parameters: eight log(SFR) ratios, plus the width of the final timebin. Following \citet{leja19b}, we place Student-t priors on the log(SFR) ratio between adjacent timebins. We center the priors such that they follow the predicted SFH for a massive quiescent galaxy at similar redshift from UniverseMachine \citep{behroozi19}. This prior is more physically-motivated than a prior centered at zero, which would imply that galaxies form stars at a constant rate across time. A forthcoming paper, Suess et al. in prep., describes this SFH model in more detail and presents mock recovery tests and comparisons with multiple other SFH parameterizations. The SFH we use here was designed to use a relatively small number of parameters to capture both a recent burst of arbitrary mass fraction and length, a variable quenching timescale, and a variable amount of residual star formation after the burst ends. Suess et al. in prep. shows that for \squiggle-like galaxies this fitting methodology recovers \tq with just 0.06~dex of scatter and \fburst with 0.12~dex of scatter.  

We define a ``burst" for each galaxy based on the output SFH. We interpolate the SFH to a uniform 1~Myr time scale, then define the burst start and end as the time when the time derivative of the SFH rises above and drops below a threshold value. This threshhold is the same for all galaxies in the sample, and was tuned by visual examination of representative SFHs. 
In addition to basic quantities such as stellar mass and A$_v$, we also report several quantities derived from this burst. We define \tq as the time when the recent starburst ended based on our SFH derivative threshold; this quantity tells us how long the galaxy has been quenched. We also define $m_{\rm{burst}}$ to be the stellar mass formed in the burst, and \fburst to be the {\it fraction} of the total stellar mass formed during the burst. Suess et al. in prep. describes these definitions in more detail, and explores several alternate definitions of \tq including the sSFR-based definitions used by \citet{tacchella21}. While the exact numerical value of \tq depends on the definition used, in general different definitions produce quantitatively similar results.

\subsubsection{Data included in the SED fits}
We fit the \texttt{Prospector} model described above jointly to both photometry and spectroscopy. We include the SDSS $ugriz$ photometry and the WISE 3.4$\mu$m and 4.6$\mu$m photometric points. However, we do not include the WISE 12$\mu$m and 24$\mu$m photometry in our fit. 
\citet{alatalo17} shows that post-starburst galaxies in the local universe have complex and unusual mid-infrared properties: the WISE photometry of these galaxies cannot be reproduced by starlight alone, but appear to be significantly influenced by emission from AGN, PAH features, and/or AGB dust. Due to the complex and still poorly understood nature of this part of the spectrum, in particular for post-starburst galaxies, the mid-IR data points will not help constrain the properties we aim to address in this paper. We ensured that excluding these points did not bias our recovered SFRs: for a randomly-selected subsample of $\sim100$ galaxies, there was no significant difference in the median SFR in fits where we included or excluded the WISE 12$\mu$m and 24$\mu$m photometry. Further analysis of the mid-IR data points are beyond the scope of this paper and will be subject to a future investigation. 
We discuss the WISE properties of our sample further in Section~\ref{sec:wise}.

\begin{figure*}
    \centering
    \includegraphics[width=.9\linewidth]{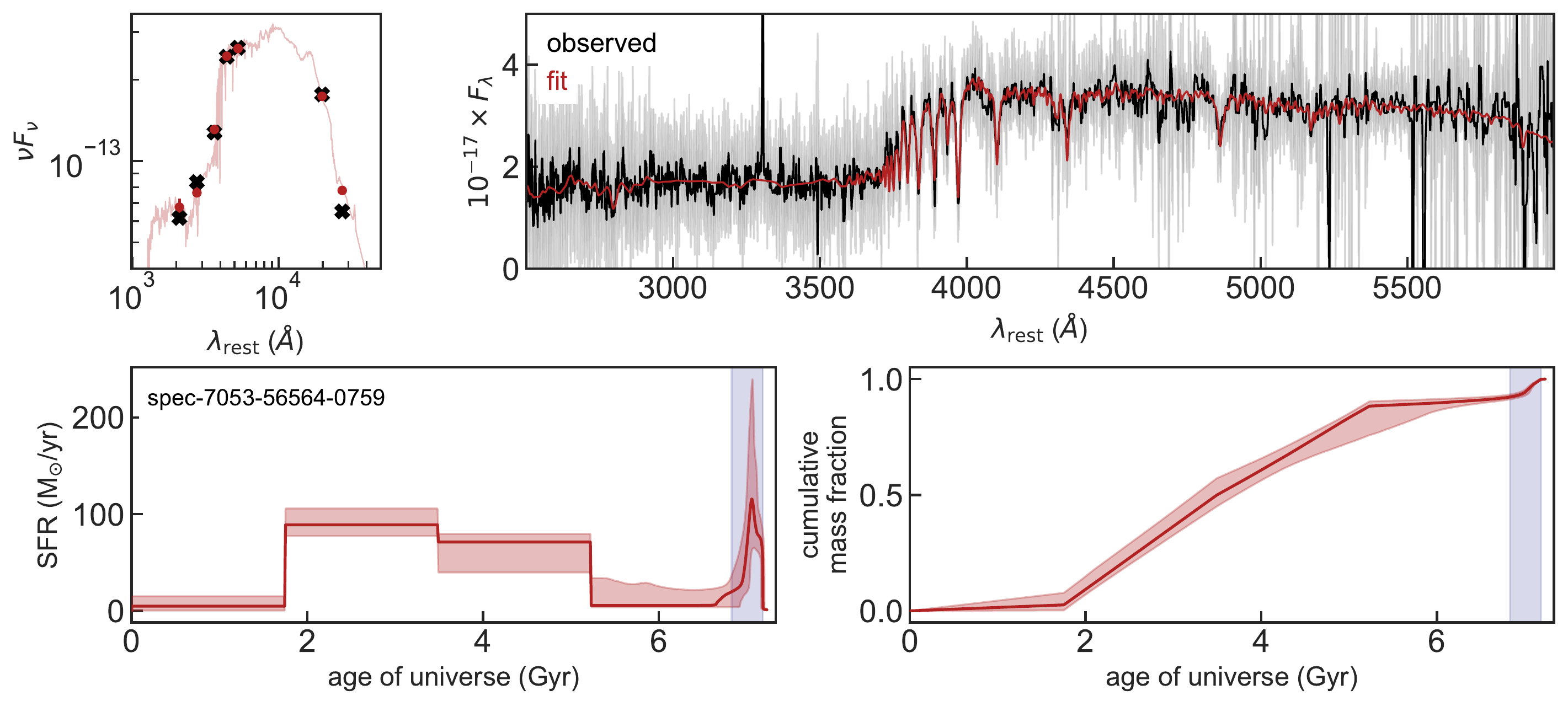}
    \includegraphics[width=.9\linewidth]{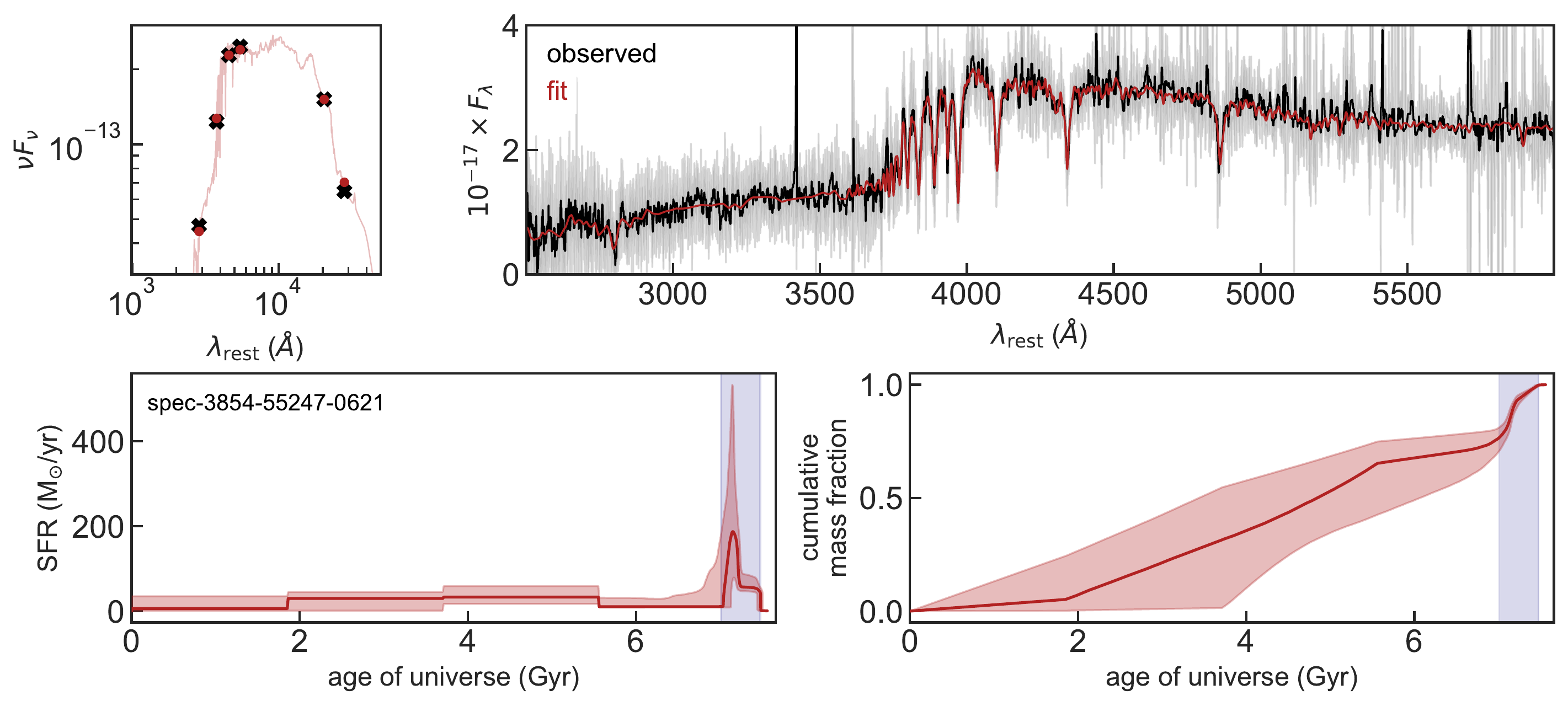}
    \includegraphics[width=.9\linewidth]{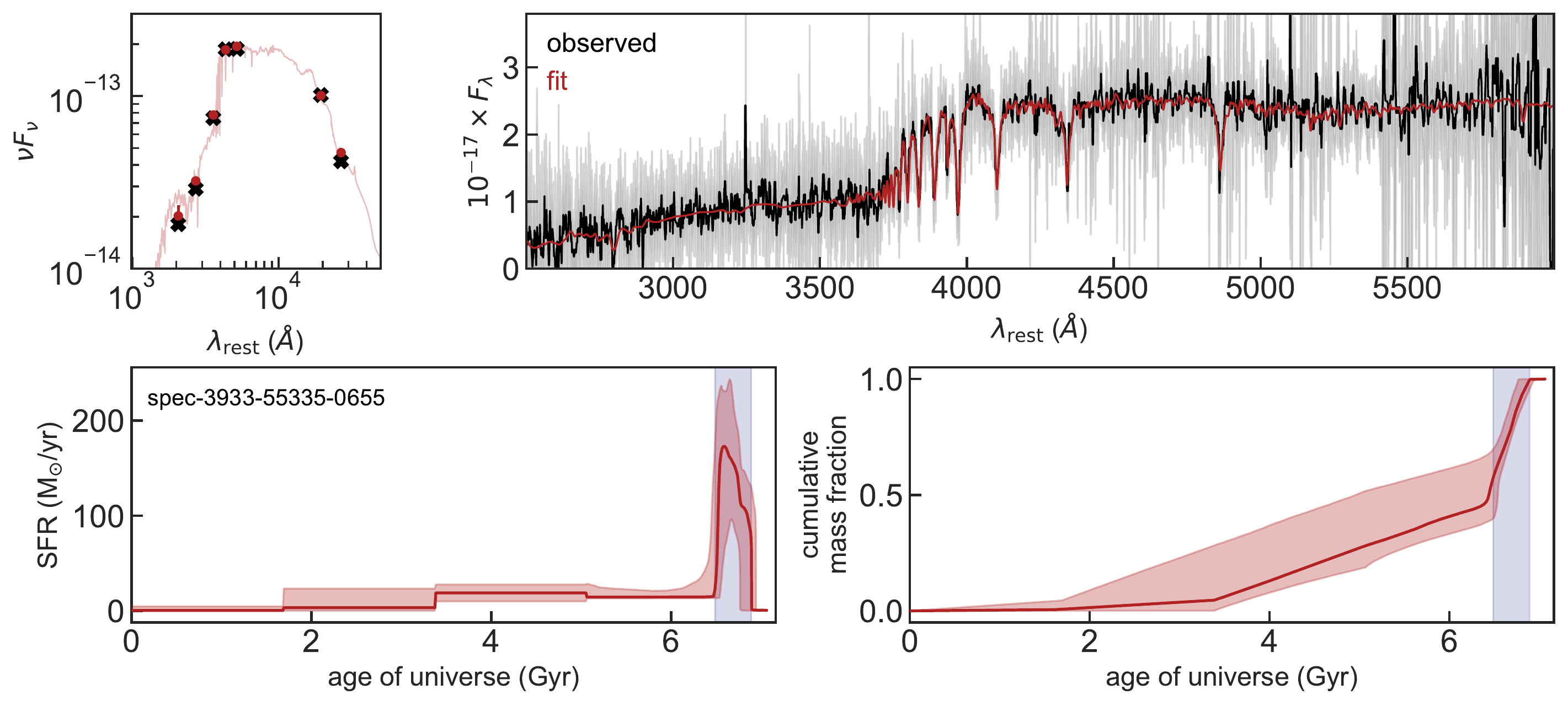}
    \caption{Three example fits using our \texttt{Prospector} model, chosen to roughly span the observed range in \tq and \fburst. The fits are ordered by increasing \tq. The upper left panel shows the observed (black cross) and best-fit (red point) SDSS and WISE photometry. The upper right panel shows the observed spectrum (grey), a 7-pixel median boxcar smoothed version of the observed spectrum (black), and the best-fit spectrum (red). The lower left panel shows the derived SFH and its 16-84th percentile confidence interval; the blue shaded region represents the ``burst" defined using a derivative-based threshhold described in the text. The lower right panel shows the cumulative fraction of the mass formed over time; again, the burst is shaded in blue.}
    \label{fig:fit_examples}
\end{figure*}

In addition to seven-band photometry, we fit the SDSS spectrum of each galaxy. We aperture-correct the spectrum using the observed SDSS photometry in the $gri$ bands. We note that at $z\sim0.7$, the diameter of the SDSS/BOSS fiber is larger than the effective radius of these compact \psbs and thus the SDSS spectrum includes the majority of the light from each galaxy; our median aperture correction is a factor of 1.3. \citet{maltby18}, \citet{suess20}, and \citet{setton20} find that \psbs have flat color gradients across similar spatial scales as the SDSS fiber, indicating that the aperture-corrected SDSS spectra are likely accurate reflections of the total integrated stellar light from these galaxies.  

In our fits, we mask all spectral pixels within $50$\AA of the 3727\AA \oii line, or within $100$\AA of the $5007$\AA \oiii line. Previous studies have found that \oii emission in post-starburst galaxies is primarily caused by LINER or AGN activity \citep[e.g.,][]{lemaux10,yan06}. \citet{greene20} shows that some \squiggle galaxies have extreme \oiii equivalent widths, again due to the presence of AGN. Because it is currently not possible to model the AGN contribution to these forbidden lines in \texttt{Prospector}, we mask them in our fits. Unobscured AGN are too blue to fall into our color-based selection algorithm (Section~\ref{sec:selection}). While \citet{greene20} does find an elevated occurance rate of obscured AGN in our sample, the continuum emission of all but the very most luminous obscured AGN are dominated by galaxy light. Masking \oii and \oiii thus ensures that our fitting results are not dominated by AGN emission.

\subsection{SED fitting results}

\begin{table*}[ht]
\centering
\caption{\texttt{Prospector} SED fitting results for \squiggle \psbs}
\label{table:fit_results}
\begin{threeparttable}
\begin{tabular}{ccccccc}
\hline \hline
SDSS ID & $\log{\rm{M}_*/\rm{M}_\odot}$ & $\log{Z/Z_\odot}$ & $A_v$ (mag)\tnote{a} & dust index\tnote{a} & \tq (Myr) & \fburst \\ \hline
spec-6137-56270-0195 & $11.40_{-0.11}^{+0.07}$ & $0.55_{-0.10}^{+0.09}$ & $0.86_{-0.09}^{+0.15}$ & $0.20_{-0.10}^{+0.07}$ & $106_{-26}^{+17}$ & $0.36_{-0.10}^{+0.24}$ \\
spec-0978-52431-0077 & $11.31_{-0.03}^{+0.06}$ & $0.20_{-0.07}^{+0.32}$ & $0.44_{-0.05}^{+0.07}$ & $-0.00_{-0.14}^{+0.19}$ & $165_{-52}^{+9}$ & $0.42_{-0.13}^{+0.07}$ \\
spec-5192-56066-0419 & $11.44_{-0.06}^{+0.06}$ & $-0.02_{-0.11}^{+0.10}$ & $0.34_{-0.06}^{+0.07}$ & $-0.69_{-0.18}^{+0.23}$ & $495_{-126}^{+123}$ & $0.51_{-0.16}^{+0.28}$ \\
spec-5288-55865-0858 & $11.05_{-0.08}^{+0.07}$ & $0.28_{-0.07}^{+0.10}$ & $0.58_{-0.10}^{+0.10}$ & $-0.21_{-0.15}^{+0.17}$ & $183_{-24}^{+21}$ & $0.38_{-0.13}^{+0.12}$ \\
spec-4575-55590-0605 & $11.30_{-0.05}^{+0.04}$ & $0.06_{-0.06}^{+0.06}$ & $0.37_{-0.07}^{+0.07}$ & $-0.03_{-0.17}^{+0.18}$ & $480_{-115}^{+88}$ & $0.41_{-0.08}^{+0.15}$ \\
spec-3817-55277-0279 & $11.59_{-0.03}^{+0.03}$ & $-0.28_{-0.08}^{+0.09}$ & $0.17_{-0.03}^{+0.03}$ & $-0.71_{-0.18}^{+0.23}$ & $715_{-217}^{+521}$ & $0.98_{-0.23}^{+0.02}$ \\
spec-5140-55836-0177 & $11.23_{-0.03}^{+0.04}$ & $0.24_{-0.11}^{+0.18}$ & $0.23_{-0.06}^{+0.05}$ & $-0.45_{-0.27}^{+0.30}$ & $217_{-81}^{+102}$ & $0.70_{-0.19}^{+0.13}$ \\
spec-1630-54476-0502 & $11.56_{-0.08}^{+0.04}$ & $0.12_{-0.07}^{+0.09}$ & $0.35_{-0.07}^{+0.06}$ & $0.22_{-0.19}^{+0.12}$ & $55_{-13}^{+14}$ & $0.15_{-0.03}^{+0.23}$ \\
spec-3754-55488-0041 & $11.54_{-0.05}^{+0.03}$ & $0.21_{-0.11}^{+0.10}$ & $0.15_{-0.06}^{+0.07}$ & $-0.48_{-0.36}^{+0.42}$ & $358_{-90}^{+105}$ & $0.16_{-0.04}^{+0.13}$ \\
spec-6649-56364-0311 & $11.47_{-0.09}^{+0.04}$ & $-0.01_{-0.17}^{+0.18}$ & $0.17_{-0.07}^{+0.05}$ & $-0.44_{-0.29}^{+0.31}$ & $352_{-115}^{+115}$ & $0.23_{-0.09}^{+0.34}$ \\
spec-5048-56218-0165 & $11.53_{-0.05}^{+0.04}$ & $0.03_{-0.09}^{+0.08}$ & $0.52_{-0.07}^{+0.09}$ & $-0.44_{-0.23}^{+0.24}$ & $271_{-94}^{+139}$ & $0.32_{-0.10}^{+0.17}$ \\
spec-6054-56089-0547 & $11.65_{-0.03}^{+0.02}$ & $0.71_{-0.10}^{+0.09}$ & $0.14_{-0.04}^{+0.04}$ & $-0.18_{-0.21}^{+0.25}$ & $94_{-22}^{+34}$ & $0.19_{-0.03}^{+0.03}$ \\
spec-4403-55536-0765 & $11.50_{-0.04}^{+0.05}$ & $0.15_{-0.08}^{+0.13}$ & $1.24_{-0.08}^{+0.07}$ & $0.11_{-0.09}^{+0.09}$ & $38_{-14}^{+20}$ & $0.56_{-0.11}^{+0.13}$ \\
spec-6032-56067-0159 & $11.16_{-0.05}^{+0.04}$ & $0.37_{-0.21}^{+0.19}$ & $0.28_{-0.08}^{+0.08}$ & $-0.33_{-0.34}^{+0.22}$ & $147_{-51}^{+52}$ & $0.32_{-0.08}^{+0.11}$ \\
spec-6639-56385-0597 & $11.54_{-0.03}^{+0.02}$ & $0.00_{-0.08}^{+0.12}$ & $0.14_{-0.04}^{+0.05}$ & $-0.75_{-0.18}^{+0.45}$ & $159_{-32}^{+44}$ & $0.47_{-0.08}^{+0.20}$ \\
spec-4013-55629-0073 & $11.48_{-0.06}^{+0.05}$ & $0.15_{-0.11}^{+0.16}$ & $0.08_{-0.04}^{+0.06}$ & $-0.21_{-0.41}^{+0.32}$ & $150_{-48}^{+77}$ & $0.32_{-0.09}^{+0.39}$ \\
spec-5993-56070-0251 & $11.46_{-0.08}^{+0.06}$ & $0.24_{-0.10}^{+0.15}$ & $0.30_{-0.10}^{+0.13}$ & $-0.40_{-0.29}^{+0.32}$ & $130_{-37}^{+58}$ & $0.33_{-0.11}^{+0.29}$ \\
spec-5014-55717-0745 & $11.32_{-0.07}^{+0.07}$ & $0.04_{-0.13}^{+0.14}$ & $0.54_{-0.12}^{+0.12}$ & $0.13_{-0.30}^{+0.19}$ & $195_{-73}^{+179}$ & $0.43_{-0.21}^{+0.35}$ \\
spec-5291-55947-0601 & $11.30_{-0.06}^{+0.10}$ & $0.13_{-0.07}^{+0.08}$ & $0.26_{-0.09}^{+0.07}$ & $-0.38_{-0.31}^{+0.24}$ & $47_{-15}^{+34}$ & $0.44_{-0.34}^{+0.49}$ \\
spec-5475-56011-0379 & $11.60_{-0.07}^{+0.04}$ & $0.05_{-0.05}^{+0.05}$ & $0.37_{-0.06}^{+0.08}$ & $-0.34_{-0.15}^{+0.16}$ & $76_{-26}^{+36}$ & $0.33_{-0.14}^{+0.61}$ \\
... & ... & ... & ... & ... & ... & ... \\ \hline
\end{tabular}
\begin{tablenotes}
\item[a]{Assuming the \citet{kriek10} dust law.}
\item (This table is available in its entirety in a machine-readable form in the online journal. A portion is shown here for guidance regarding its form and content.)
\end{tablenotes}
\end{threeparttable}
\end{table*}

Here, we show the results of our SED fitting to the full \squiggle sample.
Figure~\ref{fig:fit_examples} shows example \texttt{Prospector} fits to three galaxies in \squiggle. The upper left panel shows the observed and best-fit SDSS and WISE photometry; the upper right panel shows the observed SDSS spectrum and the best-fit stellar population model; the lower left panel shows the derived SFH; as an alternate way of viewing the SFH, the lower right panel shows the cumulative mass fraction formed as a function of time. The shaded vertical blue region in the third and fourth columns marks the recent starburst (using the derivative-based method described in detail in Suess et al. in prep.). This figure illustrates that our data and modeling framework are able to place strong constraints on the SFH for each galaxy.

\begin{figure}
    \centering
    \includegraphics[width=\linewidth]{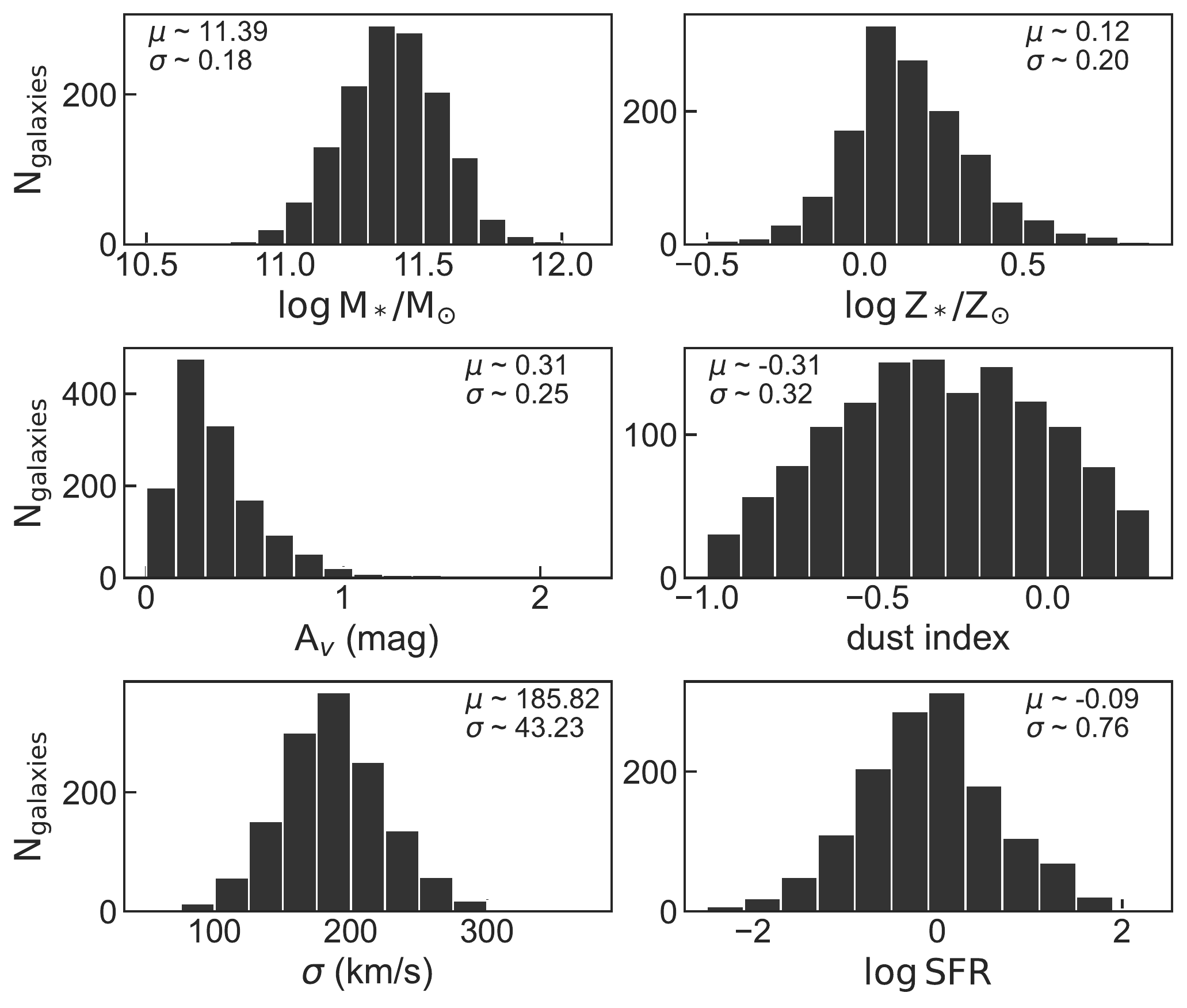}
    \caption{Histograms of recovered properties from SED fitting. The median and standard deviation of each parameter is shown in the upper left or right corner of the plot. \squiggle galaxies tend to have very high stellar masses and relatively low dust attenuation.} 
    \label{fig:prospect_hists}
\end{figure}

Figure~\ref{fig:prospect_hists} shows histograms of derived properties for the full \squiggle sample. These fits confirm that \squiggle galaxies are massive; nearly all galaxies in the sample have $\rm{M}_* \gtrsim 10^{11}\rm{M}_\odot$. This is not unexpected: these $z>0.5$ galaxies had to be bright enough to be targeted spectroscopically by SDSS and meet our S/N$\ge6$ criterion. While lower-mass \psbs may exist at $z\sim0.7$, they would not be included in our sample; \squiggle was designed to select bright, massive, and burst-dominated galaxies that can serve as testbeds for the fast quenching process. \squiggle galaxies are likely the most extreme examples of what may be a much larger population of quenching galaxies. We also see that--- as expected--- the majority of \squiggle galaxies are relatively dust-free, with a median A$_v \sim 0.3$~mag. Very dusty \psbs are generally too red to fall into our color-based sample selection (Section~\ref{sec:selection}; Figure~\ref{fig:selection_correlations}). The best-fit metallicities tend to be slightly supersolar, as expected for massive galaxies according to the \citet{gallazzi05} mass-metallicity prior we used in the fits. We see a wide range of dust indices, roughly spanning the prior range allowed in our fits. This is likely because our data have relatively little leverage on this parameter due to our short wavelength range and lack of UV and IR data. These massive galaxies have relatively high velocity dispersions, $\sigma\sim200$~km/s, as expected from the pPXF results we used as a prior \citep{greene20}.

\subsubsection{SED SFRs}

Next, we use the \texttt{Prospector} fitting results to investigate where the \squiggle sample lies in relation to the star-forming main sequence: is star formation truly suppressed in these \psbs? SED-based SFRs are notoriously tricky to calculate, and depend sensitively on the assumed SFH \citep[e.g.,][]{lee09,lee10,maraston10,wuyts11}. It is therefore {\it essential} that we test our SFH model and fitting framework before relying on SED SFRs. A forthcoming paper, Suess et al. in prep., fits the mock galaxies described in Section~\ref{sec:selection_explain} and investigates how well the output SFHs capture the ongoing SFRs of the mock \psbs. We find that SFRs above $1\rm{M}_\odot$/yr are slightly underestimated ($\lesssim0.1$~dex median offset) but generally recovered well, with $\sim0.15$ dex of scatter. However, below $1\rm{M}_\odot$/yr the data do not have significant constraining power on the SFR. 
Given the high stellar masses of our galaxies, this limit corresponds to  very low sSFRs of $\lesssim 10^{-11}\rm{\ yr}^{-1}$. Higher-quality spectra and/or additional wavelength coverage would likely be required to recover lower levels of ongoing star formation.

\begin{figure}
    \centering
    \includegraphics[width=.47\textwidth]{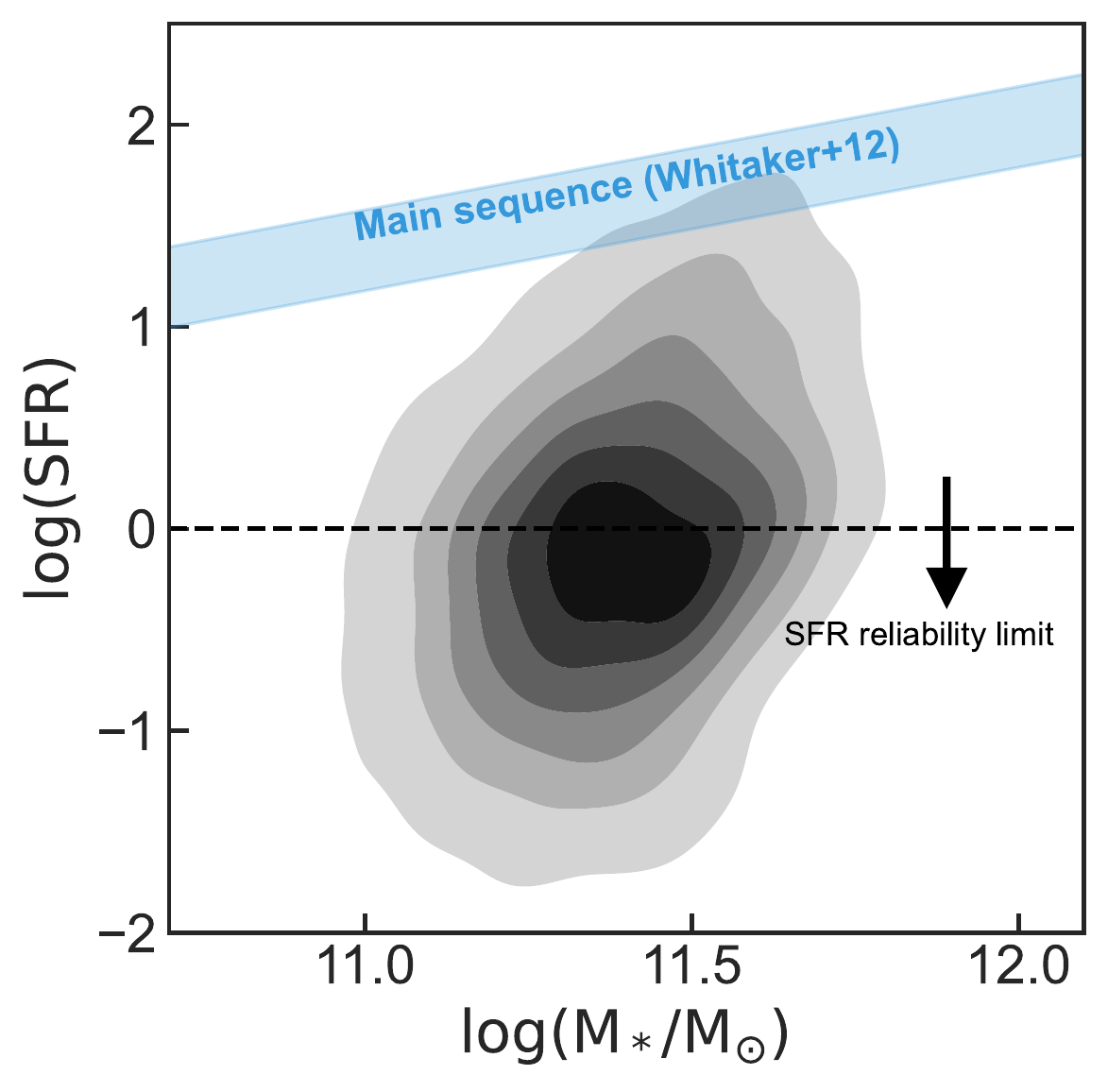}
    \caption{Star formation rate as a function of stellar mass for \squiggle galaxies (contours), compared to the $z=0.7$ star-forming main sequence from \citet{whitaker12} (blue shaded region). The contours are evenly spaced between 10\% and 85\% of the full distribution of best-fit SFRs. All recovered SFRs below $1\rm{M}_\odot/yr$ should be treated as upper limits at $1\rm{M}_\odot/yr$; our spectra and fitting methods cannot distinguish between lower levels of ongoing star formation. Galaxies in \squiggle are quenched, with the majority of galaxies lying more than an order of magnitude below the main sequence. Just 2\% of \psbs in our color-selected sample have SFRs on or above the main sequence.}
    \label{fig:SFR-M}
\end{figure}

Figure~\ref{fig:SFR-M} shows the derived SFRs as a function of stellar mass for all galaxies in the \squiggle sample. The shaded blue bar indicates the star-forming main sequence from \citet{whitaker12}. The dashed black line shows the reliability limit of our SFR measurements. 
The median SFR of the \squiggle galaxies is {\it below} our detectability threshold: the majority of \squiggle galaxies have $1\rm{M}_\odot$/yr or less of ongoing star formation. This is more than an order of magnitude offset from the star-forming main sequence, which lies at $\sim40\rm{M}_\odot$/yr at this mass and redshift. While several galaxies in our sample do have higher ongoing SFRs, these galaxies are rare: just $\sim2\%$ of \squiggle galaxies have SFRs on or above the star-forming main sequence.

We note that these SFRs are based on photometry with $\lambda_{\rm{rest}}\lesssim12\mu\rm{m}$, and thus we cannot exclude the possibility that these objects host some amount of additional star formation that is fully obscured by dust. This would require optically thick dust: otherwise, dusty star-forming galaxies have $B_m - V_m$ colors that are too red to fall into our selection algorithm. Observations at rest-frame infrared or longer wavelengths, such as those described in Section~\ref{sec:objectives}, are required to fully rule out higher obscured SFRs (additionally see, e.g., \citealt{alatalo17} and \citealt{smercina18} for a discussion of the mid- and far-infrared properties of K+A galaxies). While in Section~\ref{sec:wise} we show that most of our sample is undetected in W4, these upper limits do not allow us to eliminate the possibility that some \squiggle galaxies could be highly dust-obscured star-formers.

\subsubsection{Properties of the recent burst}

The flexibility of our SFH model allows us to characterize the recent starburst. As described in detail in Section~\ref{sec:prospector_setup}, we define the burst start and end based on the time derivative of the output SFH. Our modeling framework is able to accurately infer the end time of the recent burst (\tq) with a scatter of just 0.06~dex, as demonstrated in Suess et al. in prep. We are also able to place conservative limits on the mass fraction formed in the recent burst: while our recovered mass fractions tend to saturate at $\sim60\%$, we are able to recover lower burst mass fractions with high fidelity (scatter $\sim0.1$~dex). This saturation is likely due to the fact that our prior assumes that the SFH follows the average UniverseMachine SFH of a massive quiescent galaxy \citep{behroozi19}; this effectively asserts that these quenched galaxies likely formed a significant amount of their mass at early times. This choice allows our model to effectively ``hide" a large number of old, red stars under a large recent burst, disfavoring extremely high burst mass fractions. 

\begin{figure}
    \centering
    \includegraphics[width=.47\textwidth]{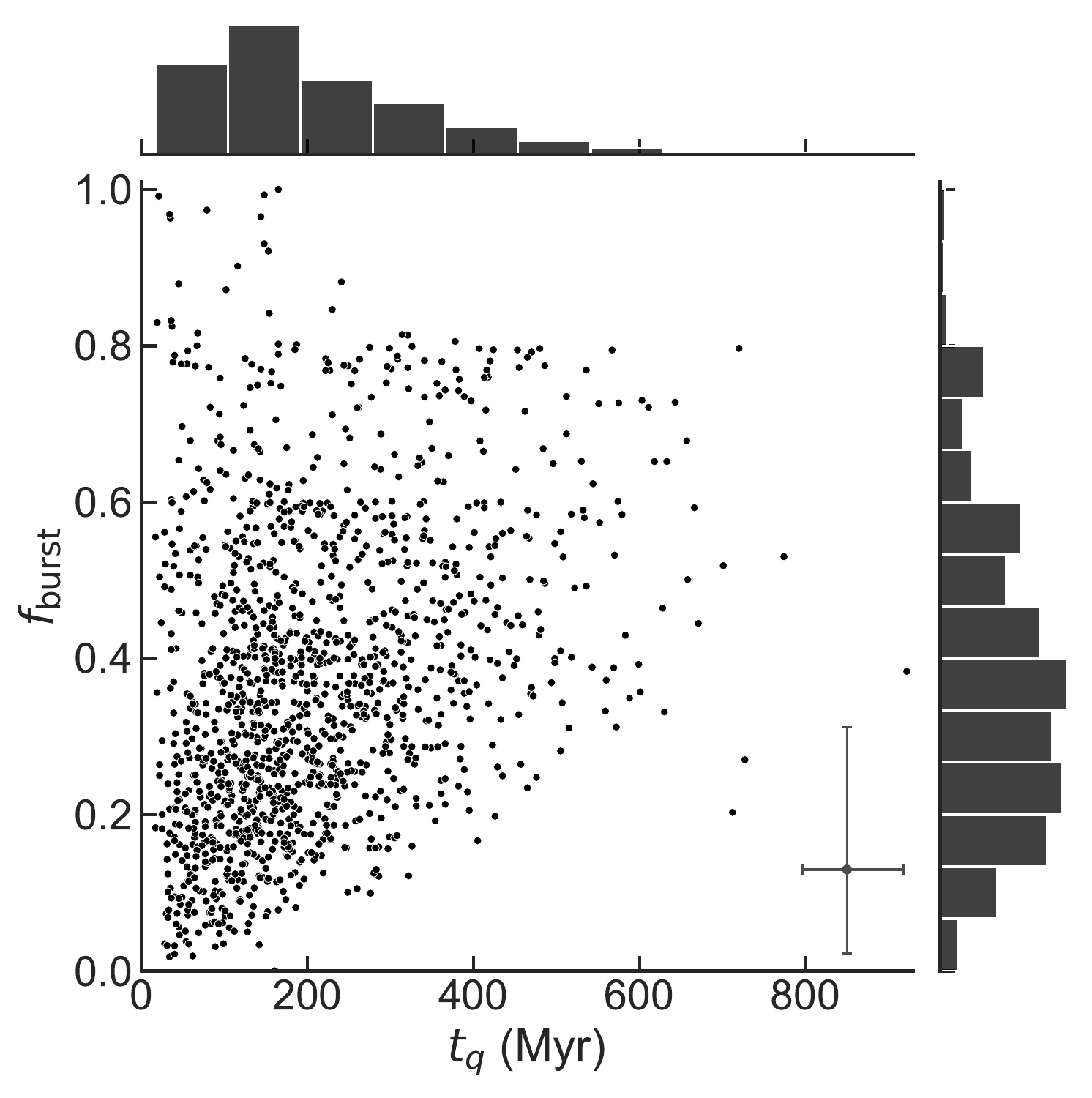}
    \caption{Burst mass fraction versus time since quenching for all \squiggle galaxies; histograms show the marginal distribution of each parameter. The point in the lower right shows a typical error bar; we note that the error on \fburst is typically asymmetric, with a higher tail towards high burst mass fractions. \squiggle galaxies have a wide range of burst mass fractions and quenching timescales. However, galaxies which quenched longer ago are only included in our sample if they also have high burst mass fractions; galaxies with high \tq and low \fburst are likely too red to satisfy our color-based selection criteria.}
    \label{fig:tq-mburst}
\end{figure}

Figure~\ref{fig:tq-mburst} shows the time since quenching (\tq) as a function of the burst mass fraction (\fburst). These SED fitting results allow us to confirm that \squiggle galaxies recently quenched a major epoch of star formation. The quenching timescale distribution peaks at $\sim175$~Myr, with a tail towards longer quenching times. Galaxies which quenched $>400$~Myr ago make up just $9\%$ of the \squiggle sample. As expected based on their spectral shapes (Figure~\ref{fig:stack_spectrum}), our post-starburst sample does not include any galaxies which quenched more than a gigayear before observation. We find that galaxies with large \tq values are included in our sample only if they also have relatively high \fburst values (as expected from Figure~\ref{fig:selection_correlations}). The 16-84th percentile range on our inferred \tq values is $\sim80-350$~Myr, which corresponds to the main sequence lifetimes of $\sim4-7\rm{M}_\odot$ stars. Thus, $\sim7\rm{M}_\odot$ stars are generally the most massive stars we expect to be alive in \squiggle galaxies. This stellar mass corresponds roughly to the B4V classification, which is the boundary where stellar spectra begin to show very deep Balmer lines and strong Balmer breaks--- exactly the spectral shapes we selected for. Our shortest recovered \tq values are thus consistent with the main sequence lifetimes of the highest-mass stars we would expect to find in these galaxies. 

Figure~\ref{fig:tq-mburst} also illustrates that, while \squiggle galaxies have a large range in burst mass fractions, in general these galaxies recently concluded a major burst. 
More than $75\%$ of \squiggle galaxies formed at least a quarter of their total stellar mass during the recent burst, and 20\% of \squiggle galaxies formed {\it more than half} of their total stellar mass in the recent burst. Despite our relatively conservative priors on burst mass fraction, we find that \squiggle galaxies appear to be much more burst-dominated than local \psbs; \citet{french18} found that $z\sim0$ \psbs often formed just $\sim10-15\%$ of their mass during the recent starburst. By going to intermediate redshift, we were able to identify the tail end of the peak quenching era at cosmic noon.

\subsection{\squiggle galaxies have anomalous mid-infrared properties}
\label{sec:wise}

\begin{figure*}[ht]
    \centering
    \includegraphics[width=.98\textwidth]{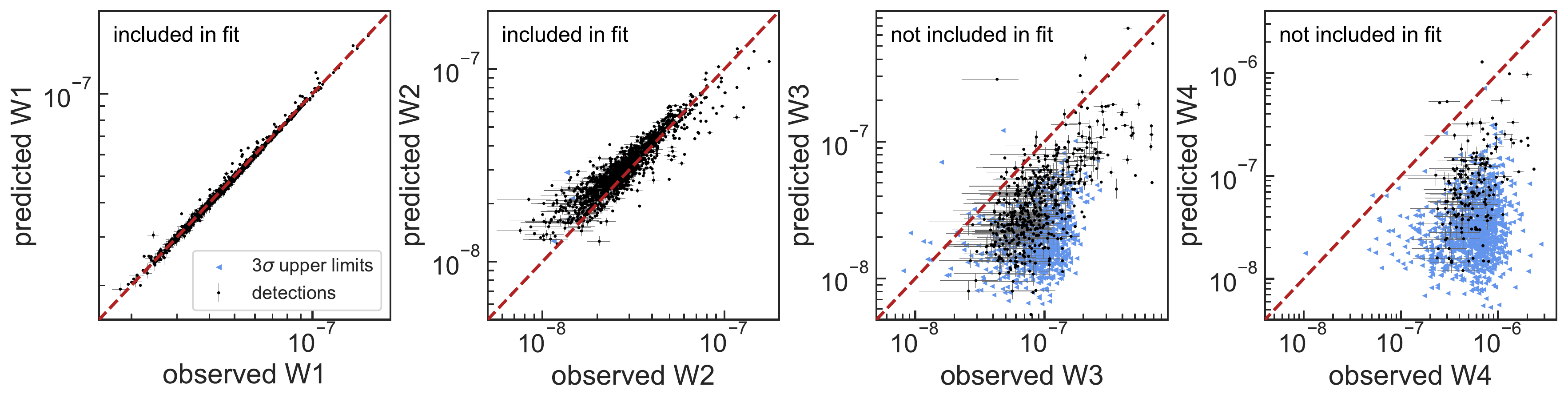}
    \caption{Predicted vs. observed WISE fluxes. Non-detections are indicated with blue triangles. While both W1 (3.4$\mu$m) and W2 (4.6$\mu$m) are well-predicted by our \texttt{Prospector} modeling, we find that \squiggle galaxies detected in the W3 (12$\mu$m) and W4 (22$\mu$m) bands tend to have much higher fluxes than predicted.}
    \label{fig:wise}
\end{figure*}

In the local universe, several studies have found that post-starburst galaxies have unusual mid-infrared properties. \citet{alatalo17} showed that $z<0.3$ post-starburst galaxies have strong infrared emission, shallow [3.4]~-~[4.6] colors, and flat or rising slopes at [12]~-~[22]. These properties are unlikely to be produced through star formation alone, and indicate the presence of AGN, strong PAH features, and/or significant contributions from dust-enshrouded AGB stars. \citet{smercina18} showed that $z\sim0$ post-starburst galaxies exhibit high PAH abundances and very strong PAH features as well as large reservoirs of warm dust indicating atypical radiation fields. These features are unusual for either star-forming or quiescent galaxies, and indicate that a ``standard" SED fitting setup is unlikely to accurately reproduce the mid-infrared properties of post-starburst galaxies. 

These results lead us to question whether the \psbs in \squiggle may have unusually high WISE fluxes or anomalous WISE colors. Because the exact nature of the processes contributing to this mid-IR flux is uncertain, we did {\it not} include the WISE 12$\mu$m (W3) or 22$\mu$m (W4) photometry in our SED fitting. Here, we test whether the {predicted} WISE photometry from our \texttt{Prospector} modeling matches the observed data points.

We first carefully flag which \squiggle galaxies are detected in the W3 and W4 bands. We consider all W3 and W4 measurements $\ge3\sigma$ to be detections. Because W1 and W2 are both significantly more sensitive than W3/W4 and have much better characterized PSFs, we then use the positions of W1 and W2 sources to confirm lower-S/N W3 and W4 sources. We consider W3 and W4 measurements at $>1.7\sigma$ ($>90\%$ confidence) to be detections if they correspond to a W1/W2 source detected at $>2.5\sigma$. While individually these lower-S/N sources would not be formal detections, the combination of multiple bands of WISE imaging results in a robust $>3\sigma$ identification of a source, albeit with large uncertainties on the individual WISE fluxes. For all sources that do not meet these criteria, we set robust $3\sigma$ upper limits on the W3 and W4 flux based on the measured background in the WISE bands. In total, we find that 511 \squiggle galaxies are detected in W3 and 176 are detected in W4.

Figure~\ref{fig:wise} shows the WISE fluxes predicted from our \texttt{Prospector} modeling as a function of the observed WISE fluxes. The left two panels show W1 (3.4$\mu$m) and W2 (4.6$\mu$m), both of which are included in our \texttt{Prospector} fits; the right two panels show W3 (12$\mu$m) and W4 (22$\mu$m), which are {\it not} included in our fitting. The dashed red line in each panel shows the one-to-one relation. We find that the observed W1 and W2 values are well-matched by our fits, with a $<0.005$~dex offset in W1 and a 0.03~dex offset in W2. However, the fluxes of \squiggle galaxies that are detected in W3 and W4 tend to be significantly underpredicted by our modeling. In W3, detected galaxies have fluxes $\sim0.4$~dex higher than our predictions; for W4 detections, the observations are $\sim0.8$~dex higher than the predictions. We note that including the W3 and W4 data points in our modeling does not resolve these offsets, indicating that the mid-infrared properties of these galaxies cannot be captured by our modeling framework. 
Understanding the nature of this excess mid-infrared flux is beyond the scope of this work, and will be studied in future \squiggle papers.

\section{SFRs from lines}
\label{sec:line_sfrs}
In this section, we compare the SFRs obtained by our SED fitting (Figure~\ref{fig:SFR-M}) with several alternate techniques for estimating SFRs based on optical lines; our goal here is to compare our \texttt{Prospector} SFRs with other SFR indicators commonly used in the literatures. 
Two spectral features in the SDSS wavelength regime--- H$\beta$ and \oii$\lambda$3727--- are often used as SFR indicators. 
Both of these lines are imperfect SFR indicators for post-starburst galaxies. First, due to the strong A-star signatures of our \psbs (Figure~\ref{fig:stack_spectrum}), we typically see H$\beta$ in absorption, not emission. While jointly modeling the continuum absorption along with the line emission can provide estimates of how much H$\beta$ emission is filling in the absorption line, these line flux measurements are extremely difficult for massive galaxies with broad lines. 
Second, LINER emission can contribute to both \oii and H$\beta$ line flux; previous studies have found that this LINER emission is common in post-starburst galaxies \citep[e.g.,][]{lemaux10,kocevski11,french15}. Shocks could also contribute to the ionized emission from galaxies, including \oii \citep[e.g.,][]{alatalo16,maddox18}. In this case, \oii and H$\beta$ SFRs should be treated as {\it upper} limits, as in \citet{french15}: star formation is not the only contributor to the measured line flux. However, both H$\beta$ and \oii are affected by dust. The SDSS spectra do not cover the H$\alpha$ regime, and thus our best estimate of the dust attenuation in these galaxies comes from our SED fitting. If these dust values are underestimated, the SFR values should be considered {\it lower} limits. The competing effects of dust and LINER emission thus make it unclear whether H$\beta$ and \oii SFRs should be treated as upper or lower limits for post-starburst galaxies: the answer likely depends on the individual galaxy. For this reason--- like \citet{belli21}--- we conclude that the SED SFRs are more reliable than line-based SFRs for post-starburst galaxies. Nonetheless, here we compute both H$\beta$ and \oii SFRs to serve as a comparison point for the \texttt{Prospector} SFRs calculated in Section~\ref{sec:prospector}.

\begin{figure}
    \centering
    \includegraphics[width=.47\textwidth]{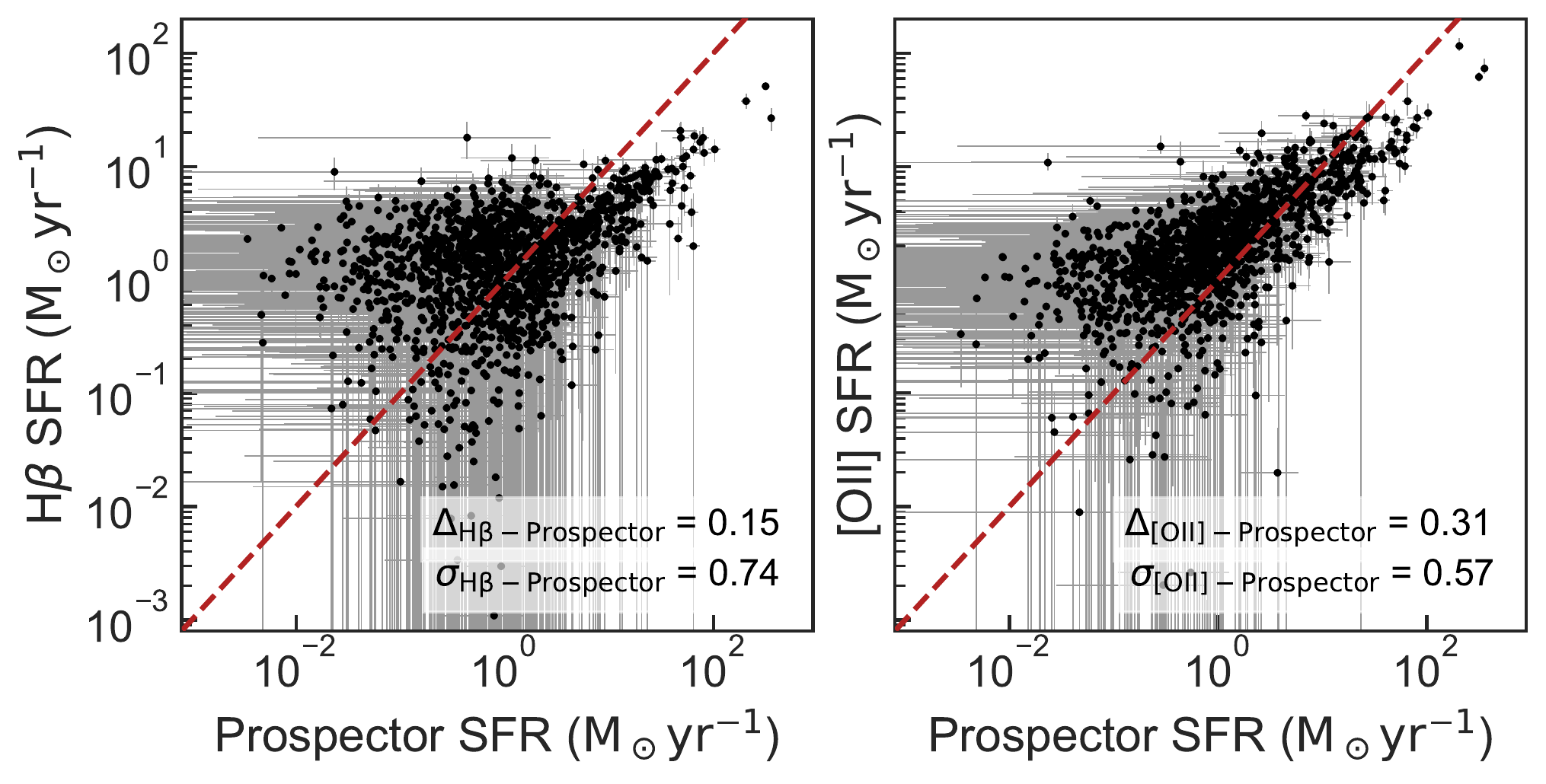}
    \caption{Comparison of SFR measured directly from both the H$\beta$ and \oii lines with the \texttt{Prospector} SFRs described in Section~\ref{sec:prospector}. There is significant scatter in both line measurements at the low-SFR end. Towards high SFRs, the \oii SFRs are offset slightly higher than the \texttt{Prospector} SFRs, likely due to LINER contributions; the H$\beta$ SFRs are offset lower than the \texttt{Prospector} SFRs.}
    \label{fig:sfr_comparison}
\end{figure}

We adopt H$\beta$ line flux measurements from \citet{greene20}, who used the public penalized pixel-fitting code pPXF \citep{cappellari04} to fit the stellar continuum as well as H$\beta$ and \oiii emission lines. We assume an intrinsic line ratio of $\rm{H}\alpha / \rm{H}\beta = 2.86$ to estimate the H$\alpha$ flux. We correct these fluxes for dust using the \texttt{Prospector} dust index and A$_v$ values. Again, our fitting assumes that lines (including H$\alpha$ and H$\beta$) are twice as attenuated as the stellar continuum. We then use the \citet{kennicutt98} conversion, adjusted to a \citet{chabrier03} IMF, to estimate the SFR. We note that our \texttt{Prospector} fits include the H$\beta$ line, and thus this line is a also a major contributor to the SFRs shown in Figure~\ref{fig:SFR-M}. However, the \citet{kennicutt98} conversion assumes solar metallicity and a constant SFH, whereas the \texttt{Prospector} SFRs takes into account the metallicity and SFH.

To calculate \oii SFRs, we measure the aperture-corrected flux from the \oii doublet by modeling the region of the spectrum around the line with a single Gaussian centered at the mean wavelength of the \oii doublet plus a straight-line fit to the continuum. Our spectra do not resolve the \oii doublet. We hold the width of the Gaussian line profile fixed to the velocity as measured by pPXF \citep{greene20}. Again, we correct the measured line flux for dust extinction using the \texttt{Prospector} fitting results. Finally, we used the conversion in \citet[][]{kennicutt98} (adjusted to a \citealt{chabrier03} IMF) to convert our measured line flux to a SFR. Error bars on the SFR were obtained using 1,000 bootstrap realizations of the measured line flux. 

Figure~\ref{fig:sfr_comparison} shows the SFR estimated from both H$\beta$ (left) and \oii (right) as a function of the \texttt{Prospector} SFR (Section~\ref{sec:prospector}). Again, our mock recovery tests indicate that \texttt{Prospector} SFRs do not have significant constraining power below $1\rm{M}_\odot\rm{yr}^{-1}$. We see that, while the line-based SFRs generally correlate with the \texttt{Prospector} SFRs, there is a large amount of scatter, particularly at the low-SFR end. H$\beta$ SFRs are typically higher than \texttt{Prospector} SFRs at low SFRs, likely because the H$\beta$ fluxes in this regime are dominated by spectral noise. At high SFR, H$\beta$ SFRs are offset lower than the \texttt{Prospector} SFRs; this could be due to the fact that the \texttt{Prospector} SFRs do not assume a constant SFH and solar metallicity, as the \citet{kennicutt98} conversion does. Median \oii SFRs are offset slightly from \texttt{Prospector} SFRs at all SFRs. This difference is largest at the lowest SFRs, likely due to an increasing fraction of \oii flux originating from LINER emission. For all three SFR indicators, there are a small fraction ($\sim2\%$) of \squiggle galaxies that have high SFRs consistent with the star-forming main sequence.  
Because of the effects of both LINER emission and dust, we take the \texttt{Prospector} SFRs as the most reliable SFR indicator available for the full \squiggle sample \citep[see also, e.g.,][]{belli21}. However, we note that no matter which SFR indicator is used--- \texttt{Prospector}, H$\beta$, or \oii--- the median SFRs of all galaxies in the \squiggle sample is an order of magnitude or more below the main sequence.

\section{Comparison to other samples of PSBs}
\label{sec:comparison}

Here, we place the \squiggle sample in the context of existing large samples of post-starburst galaxies. As discussed in Section~\ref{sec:selection}, a myriad of studies have selected recently-quenched galaxies across redshift by their spectral line strengths, spectral shapes, or colors. 
The largest spectroscopic samples of post-starburst galaxies come from the low-redshift universe, $z\lesssim0.2$, where large spectroscopic surveys such as the SDSS allow for the identification of rare recently-quenched galaxies \citep[e.g.,][]{zabludoff96,quintero04,french15,alatalo16}. While post-starburst galaxies are common at $z\gtrsim1$ \citep[e.g.,][]{whitaker12_psb,wild16,belli19}, spectroscopic samples tend to be smaller due to the difficulty of high-redshift spectroscopy. The only other large intermediate-redshift spectroscopic sample of post-starburst galaxies comes from \citet{pattarakijwanich16}, who also select from the SDSS \& BOSS surveys. In Section~\ref{sec:compare_acrossz}, we compare \squiggle to lower- and higher-redshift samples; in Section~\ref{sec:compare_samez}, we compare \squiggle to the \citet{pattarakijwanich16} sample at the same redshift.

\subsection{Comparison to post-starburst samples across redshift}
\label{sec:compare_acrossz}

\begin{figure}
    \centering
    \includegraphics[width=.49\textwidth]{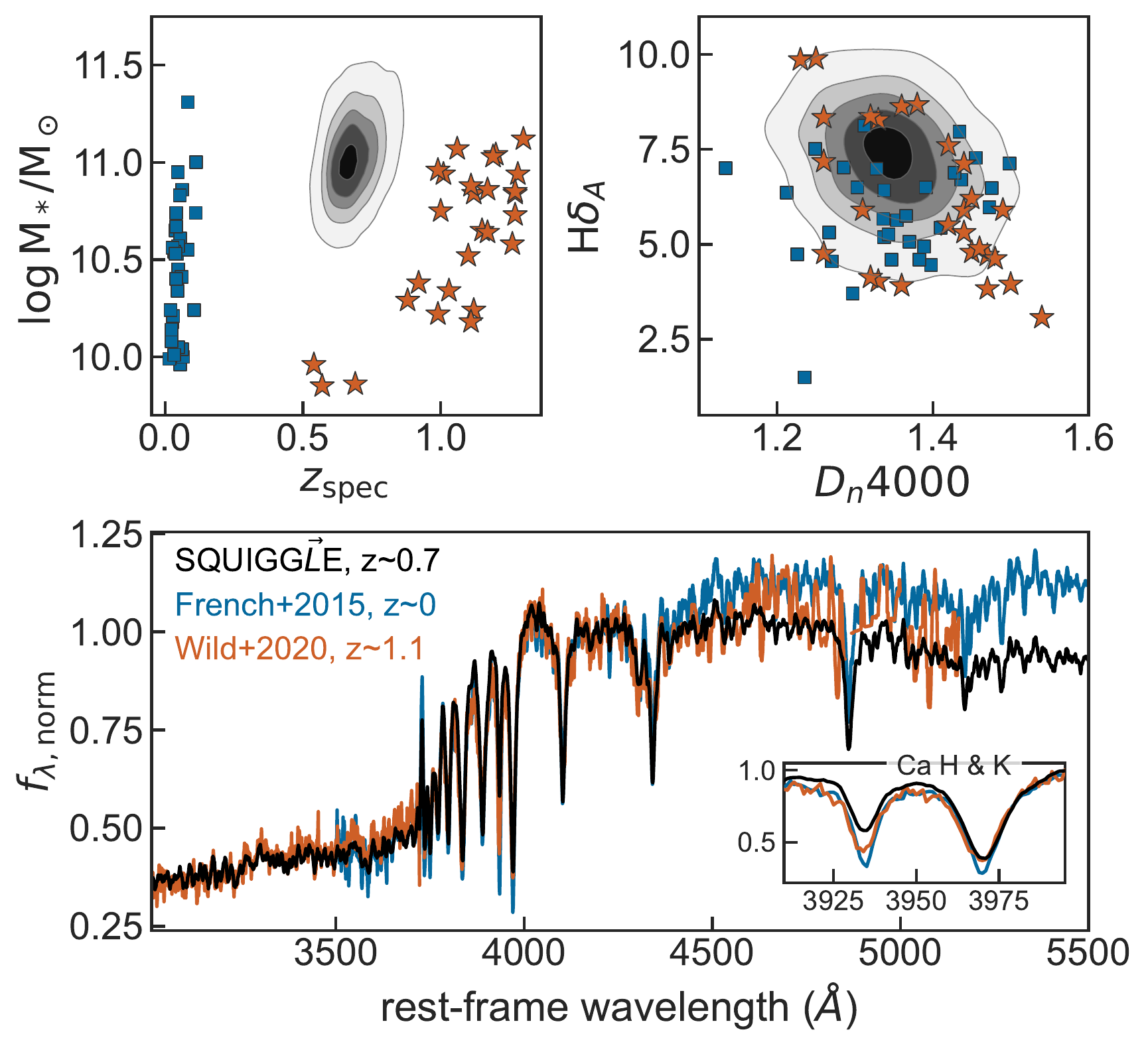}
    \caption{Comparison between the \squiggle sample (black) and other samples of post-starburst galaxies, including the \citet{french15} sample of $z\sim0$ galaxies (blue) and the higher-redshift sample from \citet{wild20} (orange). The \squiggle sample lies at high stellar mass and intermediate redshift. The bulk of \squiggle galaxies have similar $D_n4000$ values but slightly higher H$\delta$ values than either the \citet{wild20} or \citet{french15} samples, indicating slightly younger stellar ages. The \citet{french15} and \citet{wild20} stacks show slightly more flux at $\lambda_{\rm{rest}}>4500$\AA\ and more symmetric Ca H \& K features than the \squiggle stack; these differences likely reflect a higher fraction of older stars.}
    \label{fig:comparison}
\end{figure}

Figure~\ref{fig:comparison} shows the \squiggle sample in context of one low-redshift \citep{french15} and one high-redshift \citep{wild20} sample of post-starburst galaxies. The \citet{french15} sample was selected by their high H$\delta$ and low H$\alpha$ equivalent widths, while the \citet{wild20} sample was selected using a PCA technique. Figure~\ref{fig:comparison} shows that \squiggle galaxies lie at intermediate redshift, and tend to have higher stellar masses than many of the \citet{french15} or \citet{wild20} galaxies. The lack of low-mass galaxies in \squiggle compared to lower-redshift studies is likely a selection effect (e.g., Malmquist bias): the galaxies our sample {\it must} have had high mass in order to fall into the SDSS spectroscopic sample and exceed our S/N cut. While the three populations of post-starburst galaxies overlap significantly in H$\delta$ - $D_n4000$ space, the bulk of the \squiggle galaxies lie at slightly higher H$\delta_A$ values than the \citet{french15} or \citet{wild20} samples. Because stellar age typically increases towards the lower right of this diagram, this indicates that \squiggle galaxies are slightly {\it younger} on average than either of these samples. This is confirmed by SFH modeling: both \citet{wild20} and \citet{french18} found older post-burst ages for their samples than we find in Section~\ref{sec:prospector}. However, it is difficult to directly compare the inferred $t_q$ values because each study uses a slightly different SFH parameterization and definition of $t_q$. 

The bottom panel of Figure~\ref{fig:comparison} shows a stack of the spectra in all three samples. Both the \squiggle and \citet{french15} stacks are created using public SDSS data. As in Figure~\ref{fig:stack_spectrum}, all individual spectra are normalized using the flux between 4150 and 4250 \AA. Spectra from the \citet{wild20} sample come from both the UDSz ESO Large Programme (PI: Almaini) and \citet{maltby16}; Maltby et al. in prep. provides further details about the spectroscopic data reduction. In addition to normalizing the \citet{wild20} spectra (again, between 4150 and 4250 \AA), we smooth the stack using a nine-pixel median filter past $\sim4300$\AA, where fewer than 15 spectra have wavelength coverage. We do not show the \citet{wild20} stacked spectrum past $\sim5150$\AA, where fewer than five individual spectra have wavelength coverage. 

We see two notable differences in the stacked spectra shown in Figure~\ref{fig:comparison}: the depth of the absorption lines, and the spectral shape redward of $\sim4400$\AA. The calcium H \& K lines are deep and nearly symmetric for the \citet{french15} stack, while they are shallower and much more asymmetric in the \squiggle stack (because the CaH line is more significantly contaminated by H$\epsilon$). These spectral differences indicate that \squiggle galaxies are younger than the \citet{french15} galaxies. 
We also find that the \citet{french15} spectra have nearly a flat slope redward of $\sim4400$\AA, while (by construction) the \squiggle stack shows a relatively blue slope. These differences could be caused by differences in the sample selection technique, redshift evolution in the post-starburst population \citep[e.g.,][]{whitaker12_psb,wild16,rowlands18,belli19} or both. The \citet{wild20} stack has a slope intermediate between the \citet{french15} and \squiggle samples. This difference in slope indicates that both the \citet{french15} and \citet{wild20} samples contain galaxies with larger contributions from old stars. For the \citet{french15} sample, this aligns with expectations from the SFH fitting in \citet{french18}: these galaxies have best-fit burst mass fractions that are often $\lesssim20$\%, lower than the \squiggle burst mass fractions shown in Figure~\ref{fig:tq-mburst}. However, \citet{wild20} finds burst mass fractions that are often $\sim60-80\%$, above the median burst mass fractions for the \squiggle sample. This difference is likely primarily caused by differences in the fitting methodology and definition of ``burst mass fraction" used in \citet{wild20} and this paper. Furthermore, the \citet{wild20} stack shown in Figure~\ref{fig:comparison} is dominated by low-redshift, lower \fburst galaxies at these longer wavelengths.

\subsection{Comparison to the \citet{pattarakijwanich16} sample}
\label{sec:compare_samez}

\begin{figure*}[ht]
    \centering
    \includegraphics[width=.99\textwidth]{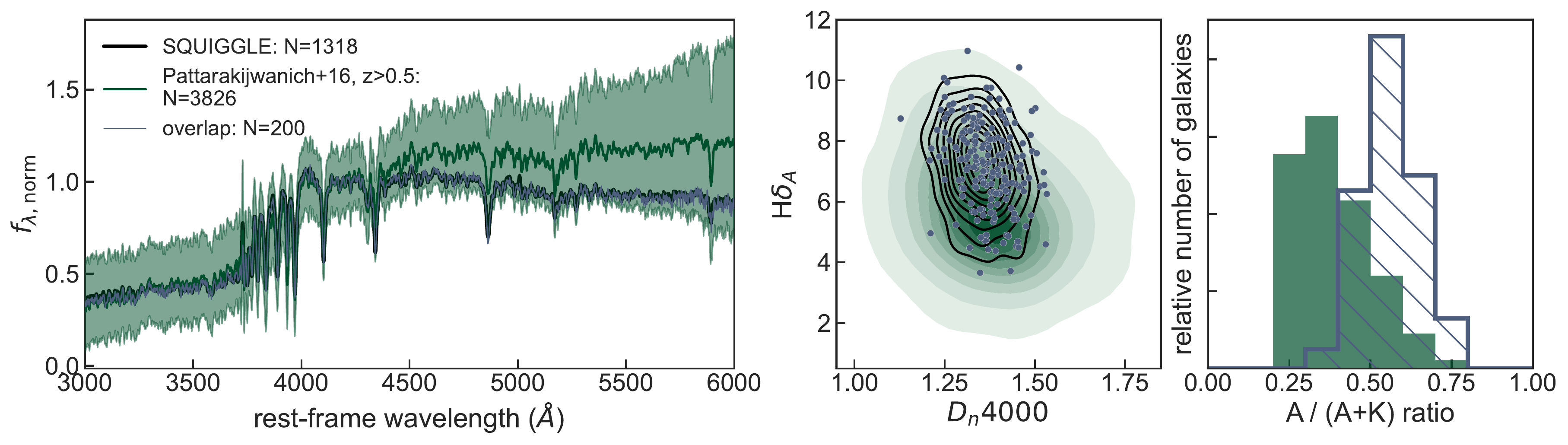}
    \caption{Comparison between the \squiggle sample (black) and the \citet{pattarakijwanich16} sample, which is also selected from the SDSS. The left panel shows a stacked spectrum of all \squiggle galaxies (black), all \citet{pattarakijwanich16} galaxies with $z\ge0.5$ (green; 16-84\% range shaded), as well as galaxies identified as post-starburst in {\it both} samples (blue; overlaps with black spectrum). The central panel shows the same three sets of \psbs in H$\delta$ and $D_n4000$, with the overlapping sample shown as blue dots. While on average the \citet{pattarakijwanich16} sample includes older galaxies with lower burst mass fractions, the distribution of this sample is very broad and the $1\sigma$ range just encompasses young \psbs like those found by \squiggle. The right panel shows histograms of the A/(A+K) ratio (e.g., the light-weighted fractional contribution of A-type stars) for the full \citet{pattarakijwanich16} as well as the overlap between the \citet{pattarakijwanich16} and \squiggle samples. The galaxies selected by \squiggle tend to have higher A/(A+K) ratios than the full \citet{pattarakijwanich16} sample.}
    \label{fig:comparison_pete}
\end{figure*}

In Figure~\ref{fig:comparison_pete}, we compare the \squiggle sample to the \citet{pattarakijwanich16} sample. This study also selected \psbs from SDSS spectroscopy. Unlike \squiggle, \citet{pattarakijwanich16} used a template-fitting approach to find \psbs: galaxies were fit with a sum of both an old (`K') component and a young (`A') component, then identified as post-starburst if the light-weighted A/(A+K) ratio exceeded 0.25. \citet{pattarakijwanich16} selected spectra from both the SDSS and BOSS surveys, including galaxies with redshifts as low as $z=0.05$. In Figure~\ref{fig:comparison_pete}, to facilitate a direct comparison with \squiggle we show only galaxies with $z\ge0.5$. The median stacked spectrum of the $z>0.5$ \citet{pattarakijwanich16} sample has significantly more flux at longer wavelengths; again, this indicates older stellar ages. However, the 16-84\% confidence interval of the \citet{pattarakijwanich16} encompasses a large range of spectral slopes. The lower 16\% interval encompasses the \squiggle stack, indicating that the \citet{pattarakijwanich16} sample includes some \psbs as young as those in \squiggle. This larger median age and wider age spread is also reflected in the $H\delta$ -- $D_n4000$ values: the median value of the \citet{pattarakijwanich16} sample is offset from the \squiggle median, but the distribution is broad and overlaps significantly with \squiggle. 

We find that 210 \psbs are selected both by \squiggle and \citet{pattarakijwanich16}. These galaxies reflect the full distribution of \squiggle galaxies: they trace the \squiggle contours in $H\delta$ -- $D_n4000$ space, and their stacked spectrum is indistinguishable from the full \squiggle sample. This relatively small number of overlapping galaxies is primarily due to the fact that \citet{pattarakijwanich16} selected galaxies from SDSS DR9, whereas the \squiggle selection is performed on SDSS DR14. Nearly half of \squiggle-identified \psbs were observed {\it after} DR9, and thus could not have been included in \citet{pattarakijwanich16}.

The right panel of Figure~\ref{fig:comparison_pete} shows the A/(A+K) ratio as calculated by \citet{pattarakijwanich16}, both for the full \citet{pattarakijwanich16} sample and the galaxies selected by both \citet{pattarakijwanich16} and \squiggle. \squiggle galaxies have a higher median A/(A+K) ratio than the full \citet{pattarakijwanich16} sample, indicating that we are generally selecting more burst-dominated galaxies.

In summary, the \squiggle sample is smaller and more targeted than the \citet{pattarakijwanich16} sample: \squiggle consists of uniformly young \psbs with a narrow spread in H$\delta$-$D_n4000$ space. The \citet{pattarakijwanich16} sample includes a wider range of stellar ages, but does not include the majority of young \squiggle galaxies which were observed after DR9.

\section{\squiggle science objectives}
\label{sec:objectives}
The large sample of bright, intermediate-redshift, recently-quenched galaxies in \squiggle enables a wide range of studies into the rapid quenching process. Most of these science cases rely on the stellar population synthesis modeling presented in this paper in combination with other multi-wavelength datasets. \squiggle is not intended to be a complete sample of \psbs at these redshifts: the selection function is complex, and not conducive to number density studies. Instead, \squiggle was designed to select the brightest, most massive, most burst-dominated \psbs at intermediate redshifts. These galaxies serve as {\it laboratories} to conduct detailed multi-wavelength dives into the processes responsible for shutting down star formation. 
Here, we briefly summarize the primary science objectives of \squiggle.

\subsection{Molecular Gas Reservoirs}
Theoretical quenching mechanisms generally rely on processes which remove the available fuel for star formation by depleting, heating, or ejecting cold molecular gas reservoirs. One of the primary science objectives of \squiggle is to directly test this assumption. An initial ALMA study of the CO(2--1) emission of two \squiggle \psbs revealed abundant gas reservoirs despite low ongoing star formation rates \citep{suess17}. This result indicates that--- contrary to expectations--- quenching does not require the total removal of molecular gas. Massive gas reservoirs have also been found in local K+A galaxies \citep[e.g.,][]{rowlands15,french15,alatalo16,smercina18} as well as several young quiescent galaxies at $1\lesssim z\lesssim1.5$ \citep{williams21, belli21}. These results have prompted theoretical studies into why star formation is suppressed in \psbs \citep[e.g.,][]{davis19,salim20}.

We are currently conducting an ALMA survey of the molecular gas content of eleven additional \squiggle \psbs (R. Bezanson et al., in prep.). This study will allow us to test whether these abundant gas reservoirs are common after quenching, and whether the gas fraction depends on other galaxy properties such as time since quenching (Section~\ref{sec:prospector}). 

These molecular gas measurements, in combination with SFRs, will allow us to place these galaxies on the Kennicutt-Schmidt relation and investigate how efficiently they are forming stars \cite[e.g.,][]{kennicutt98_ks}. As part of this work, we are in the process of obtaining other robust estimators of the SFR in these galaxies. In particular, we are using the VLA to investigate possible highly-obscured SFR, and using Keck/NIRES to calculate Balmer decrement-corrected H$\alpha$ SFRs.

\subsection{Morphologies \& Sizes}
\begin{figure}[ht]
    \centering
    \includegraphics[width=.49\textwidth]{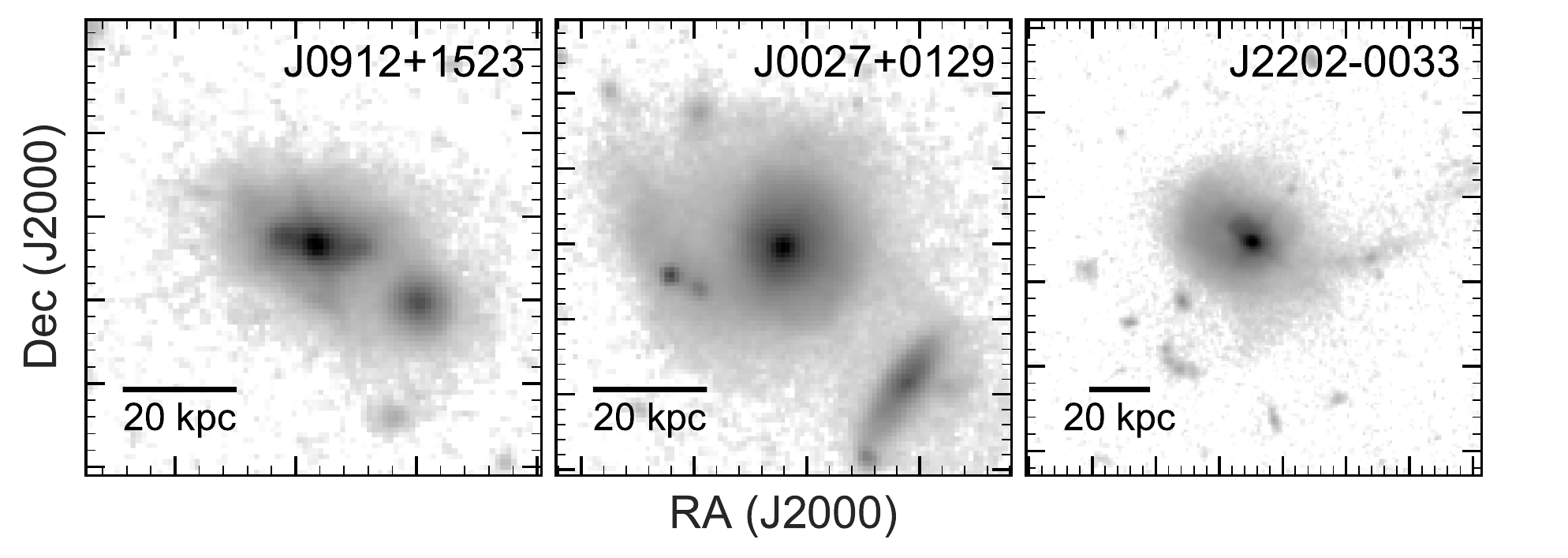}
    \caption{{\it HST}/WFC3 F110W images of three galaxies in our sample. We show a 80~kpc square cutout for J0912+1523 and J0027+0129. J2202-0033 is shown in a 150~kpc square cutout to capture the full extent of the tidal tail that stretches to the west. All three galaxies exhibit complex morphologies and merger signatures.}
    \label{fig:hst}
\end{figure}

Some studies predict that quenching can be triggered by a gas-rich major merger: in this scenario, the merger funnels gas to the galaxy's center where it is consumed in an intense starburst \citep[e.g.,][]{hopkins06,wellons15}. Deep, high-resolution imaging of \squiggle galaxies will allow us to quantify the fraction of recently-quenched galaxies that show signs of recent mergers (including tidal features and asymmetric morphologies). \citet{sazonova21} suggests that these disturbed morphologies are common for \psbs at $z\sim0$. 
Our team has obtained {\it Hubble Space Telescope} WFC3/F125W imaging for three \squiggle galaxies targeted as part of our ALMA survey (Figure~\ref{fig:hst}). These galaxies are clearly disturbed: J2202-0033 has a large tidal feature to the west, and both J0027+0129 and J0912+1523 have nearby companions which may be physically associated. The image of J0912+1523 also reveals that it has a spheroidal component embedded within a disk. 

Some previous studies have found that \psbs have extremely compact sizes, and may be smaller than their older quiescent counterparts \citep[e.g.,][]{whitaker12_psb,yano16,almaini17}. \citet{suess20} suggests that these size differences may be primarily caused by the effects of radial color gradients \citep[see also][]{maltby18,setton20}. In either case, differences in the sizes and/or color gradients of \psbs and older quiescent galaxies could provide clues both to the formation mechanisms for \psbs and the processes contributing to evolution along the quiescent sequence. Deep, high-resolution images such as those shown in Figure~\ref{fig:hst} will allow us to calculate the sizes of \squiggle \psbs and compare them to a mass-matched sample of older quiescent galaxies.

While the images presented in Figure~\ref{fig:hst} shed light on the morphologies and sizes of a few individual \squiggle galaxies, obtaining {\it HST} imaging for a statistical sample of post-starburst galaxies is prohibitively expensive. A much larger set of deep, high-resolution images comes from the overlap of the \squiggle sample and the public Hyper SuprimeCam survey \citep{aihara18}. One of our major science objectives is to use the $>150$ high-quality HSC images of \squiggle galaxies to investigate the sizes, morphologies, and merger fractions of these recently-quenched galaxies (D. Setton et al., in prep.).

\subsection{Kinematics}
In general, star-forming galaxies tend to be rotationally-supported disks, while quiescent galaxies are more likely to be kinematically hot and supported by random motions \citep[e.g.,][]{emsellem11}. However, it is still not understood why a cessation of star formation correlates with a change in kinematics. Furthermore, it is unclear whether this kinematic transition occurs before, after, or at the same time as star formation shuts down. The bright, recently-quenched \squiggle galaxies provide an ideal testbed for IFU studies to directly probe the kinematics of galaxies just after quenching. An early result for one \squiggle target (J0912+1523, also shown in Figure~\ref{fig:hst}) revealed that the galaxy is a rotating disk \citep{hunt18}. A larger followup study of five additional \squiggle galaxies showed that only $\sim$half of \psbs show clear velocity gradients, while the other half are dominated by random motions \citep{setton20}. 

\subsection{Resolved Stellar Populations}
Studying the spatially-resolved properties of \psbs could provide additional clues to the quenching process. Radial age or sSFR gradients can be used to help determine whether quenching proceeded inside-out or outside-in \citep[e.g.,][]{tacchella18,woo19}; radial metallicity gradients can help distinguish between in- and ex-situ components, important when mergers are suggested as a trigger for quenching \citep[e.g.,][]{greene15,chan16,woo19}. Our preliminary study of six \squiggle galaxies shows that they have flat age gradients as probed by H$\delta$ \citep{setton20}. In the absence of strong radial dust or metallicity gradients, these flat age gradients are consistent with the flat color gradients that \citet{maltby18} and \citet{suess20} find for \psbs. \citet{setton20} find that these six \psbs have young light-weighted ages at all radii, implying that star formation shut off uniformly throughout the galaxy. In the future, we plan to expand the sample size of \squiggle \psbs with spatially-resolved stellar population measurements; facilities like {\it JWST} would be useful to perform these studies further in the infrared, where it is easier to break the age-dust-metallicity degeneracy.

\subsection{AGN Incidence}
One popular theoretical mechanism for quenching galaxies invokes strong feedback from AGN \cite[e.g.,][]{dimatteo05,hopkins06}. This feedback, possibly induced by a major merger \citep{springel05,hopkins06,wellons15}, could heat the interstellar medium and/or drive molecular gas from the galaxy, removing the fuel for star formation \citep{alatalo15}. 
Establishing a direct causal connection between AGN activity and star formation suppression has unfortunately proven difficult, in part because AGN vary dramatically on much shorter timescales than star formation. A large number of AGN surveys have shown that AGN activity depends on star formation rate and mass \citep[e.g.,][]{hickox14}. The \squiggle sample allows us to ask how many recently-quenched galaxies show AGN activity, and test whether certain types of \psbs are more likely to host AGN. \citet{greene20} used \squiggle to show that the incidence of AGN depends strongly on $D_n4000$: recently-quenched galaxies from \squiggle are ten times more likely to host an optical AGN than a mass-matched sample of older quiescent galaxies. This hints that AGN activity is indeed correlated with the quenching process in these massive galaxies.

\subsection{IR properties}
As shown in Figure~\ref{fig:wise}, \squiggle galaxies have puzzlingly high W4 fluxes. At the redshift of \squiggle, W4 corresponds to rest-frame $\sim11-14\mu$m. Previous studies at $z\lesssim0.3$ have found that the mid-infrared spectra of post-starburst galaxies are influenced by emission from AGN, TP-AGB stars, and strong PAH features \citep[e.g.,][]{alatalo17,smercina18}. In future studies, we will perform stacking analyses of the WISE imaging for \squiggle galaxies to understand the nature of this excess W4 emission. Additionally, the mid-infrared capabilities of {\it JWST} may allow us to understand the origin of this emission. These infrared studies can be paired with studies of the molecular gas reservoirs in order to obtain a more complete picture of the interstellar medium (ISM) conditions as galaxies cease forming stars.

\section{Discussion \& Conclusions}

In this paper, we present the sample selection, stellar population properties, star formation histories, and objectives of the \squiggle survey of \psbs. We select bright, intermediate-redshift, recently-quenched galaxies from the SDSS spectroscopic sample using a simple color-based selection criterion. Using just two rest-frame color cuts, we are able to isolate 1,318 post-starburst galaxies at $0.5<z\lesssim0.9$. These galaxies all have high H$\delta$ equivalent widths, low $D_n4000$ values, and BV/A-star dominated spectra indicating young stellar ages. These recently-quenched galaxies serve as laboratories to study the processes responsible for shutting down star formation in galaxies: the signatures of the quenching process should still be imprinted on their morphologies, kinematics, and gas properties. 

We use the \texttt{Prospector} spectral energy distribution fitting code to recover the stellar population parameters and star formation histories of all \squiggle galaxies. We find that these galaxies are very massive--- nearly all \squiggle galaxies have $\rm{M}_* > 10^{11}\rm{M}_\odot$--- and have relatively low dust attenuation values. Our fitting also shows that these galaxies are indeed quenched: the median SFRs recovered from our SED fitting are more than an order of magnitude below the star-forming main sequence \citep{whitaker12}. The quenched nature of this sample is consistent with SFR estimates based on both the H$\beta$ and \oii spectral lines: while these SFRs are likely less reliable than the SED SFRs due to the competing effects of LINER emission and dust \citep[see also, e.g.,][]{belli21}, the median H$\beta$ and \oii SFRs also lie well below the main sequence. Longer-wavelength data would be required to fully rule out the possibility of highly dust-obscured star formation. While many galaxies in \squiggle host obscured (type II) AGN \citep{greene20}, because we mask the \oii and \oiii lines we do not expect our SED fitting results to be dominated by AGN emission.

Our SED fitting also allow us to quantify the properties of the recent burst. By using non-parametric SFHs, our SED fitting methodology accurately recovers both how long these galaxies have been quenched (\tq) and the fraction of the total stellar mass formed in the recent burst (\fburst). 
We find that \squiggle galaxies quenched their star formation extremely recently, with a median \tq value of just 175~Myr. Galaxies which quenched longer ago, up to $\sim800$~Myr before the time of observation, are also included in \squiggle; however, these older \psbs are only selected if they also have relatively high burst mass fractions. The \fburst distribution of \squiggle galaxies peaks at around $\sim30$\% of the total stellar mass being formed in the recent burst. We note that due to our conservative choice of priors, we likely underestimate \fburst for the most extreme and burst-dominated objects. Despite this choice, we find that 20\% of the galaxies in \squiggle formed a {\it majority} of their total stellar mass during the recent burst. 

We find that these extreme objects are younger and more burst-dominated than samples of ``K+A" or ``E+A" \psbs at $z\sim0$: \squiggle galaxies have higher median H$\delta$ equivalent width, lower $D_n4000$ values, and bluer spectral slopes than the \citet{french15} sample of local \psbs. This difference is also confirmed by the SFH fitting in \citet{french18}: many local \psbs have burst mass fractions $\lesssim20$\%, in contrast to the higher burst mass fractions we find for \squiggle galaxies. While \squiggle galaxies may have slightly lower average burst mass fractions than $z\sim1$ post-starburst galaxies from \citet{wild20}, \squiggle galaxies are on average younger than the \citet{wild20} \psbs. Together, these results indicate that our selection was able to identify the rare intermediate-redshift tail of the peak epoch of quenching at $z>1$: \squiggle galaxies recently and rapidly shut down a major burst of star formation. By targeting these bright, intermediate-redshift galaxies, \squiggle is able to strike a balance between low enough redshift that follow-up observations are feasible, and high enough redshifts that we can use these galaxies to understand how galaxies shut down their major star-forming epoch.

This large sample of recently-quenched galaxies opens a wide range of future studies. Different theoretical quenching mechanisms predict qualitatively different morphologies, age gradients, kinematics, AGN incidence, and ISM conditions. We have already begun to use this sample of galaxies to constrain the mechanisms responsible for quenching. We have found that, in contrast to theoretical predictions, quenching does {\it not} require the total removal of molecular gas \citep{suess17}. Furthermore, we have found that these recently-quenched galaxies can have a range of different kinematic structures, but tend to have flat age gradients indicating that the recent starburst was not purely centrally-concentrated \citep{hunt18,setton20}. We have also shown that AGN likely play an important role in quenching: \squiggle galaxies are more than ten times more likely to host an optical AGN than a mass-matched sample of older quiescent galaxies, and the AGN fraction is even higher in the youngest \squiggle galaxies \citep{greene20}. These studies represent just the beginning of the insights that the \squiggle sample will provide into the mechanisms responsible for transforming galaxies from disky blue star-formers to quiescent red ellipticals.

\begin{acknowledgements}
We thank the anonymous referee for a constructive report which improved the quality of this manuscript. KAS thanks Dustin Lang for kindly providing unWISE fluxes for the \squiggle catalog. RSB, JEG, DS, and DN gratefully acknowledge support from NSF-AAG\#1907697. This work was performed in part at the Aspen Center for Physics, which is supported by National Science Foundation grant PHY-1607611. KAS is partially supported by the UCSC Chancellor's Fellowship. This publication makes use of data products from the Sloan Digital Sky Survey as well as the Wide-field Infrared Survey Explorer, which is a joint project of the University of California, Los Angeles, and the Jet Propulsion Laboratory/California Institute of Technology, funded by the National Aeronautics and Space Administration. Funding for SDSS-III has been provided by the Alfred P. Sloan Foundation, the Participating Institutions, the National Science Foundation, and the U.S. Department of Energy Office of Science. The SDSS-III web site is http://www.sdss3.org/.
SDSS-III is managed by the Astrophysical Research Consortium for the Participating Institutions of the SDSS-III Collaboration including the University of Arizona, the Brazilian Participation Group, Brookhaven National Laboratory, Carnegie Mellon University, University of Florida, the French Participation Group, the German Participation Group, Harvard University, the Instituto de Astrofisica de Canarias, the Michigan State/Notre Dame/JINA Participation Group, Johns Hopkins University, Lawrence Berkeley National Laboratory, Max Planck Institute for Astrophysics, Max Planck Institute for Extraterrestrial Physics, New Mexico State University, New York University, Ohio State University, Pennsylvania State University, University of Portsmouth, Princeton University, the Spanish Participation Group, University of Tokyo, University of Utah, Vanderbilt University, University of Virginia, University of Washington, and Yale University. 
\end{acknowledgements}

\software{astropy \citep{astropy:2013,astropy:2018}, seaborne \citep{seaborn}}

\bibliographystyle{aasjournal}
\bibliography{squigglebib}

\end{document}